\documentclass[twocolumn, a4paper, 11pt,book]{archive}
\usepackage{version}
\usepackage{graphicx}

\usepackage[]{natbib}

\usepackage{txfonts}
\usepackage{booktabs}
\usepackage{multirow}

\def\lsim{\mathrel{\rlap{\lower4pt\hbox{\hskip1pt$\sim$}}
    \raise1pt\hbox{$<$}}}                % less than or approx. symbol
\def\gsim{\mathrel{\rlap{\lower4pt\hbox{\hskip1pt$\sim$}}
    \raise1pt\hbox{$>$}}}                % greater than or approx. symbol%
\usepackage{xcolor}
\usepackage{draftwatermark}
\SetWatermarkText{DRAFT}
\SetWatermarkScale{0.75}

\begin{document} 

\twocolumn[{%
 \centering
%
  %\title{}
  {\center \bf \Huge Non-thermal desorption of complex organic molecules}\\
\vspace*{0.25cm}

{\Large E. Dartois \inst{1}, %\fnmsep
%        \and
          M. Chabot \inst{2},
%         \and
         T. Id Barkach \inst{2},
%         \and
         H. Rothard \inst{3},
%         \and
         B. Aug\'e \inst{4},
%        \and
         A.N. Agnihotri  \inst{5},
%         \and
        A. Domaracka \inst{3},
%         \and
         P. Boduch \inst{3}
         }\\
\vspace*{0.25cm}

$^1$   Institut des Sciences Mol\'eculaires d'Orsay (ISMO), UMR8214, CNRS - Universit\'e de Paris-Sud, Universit\'e Paris-Saclay,\\
B\^at 520, Rue Andr\'e Rivi\`ere, F-91405 Orsay, France\\
              \email{emmanuel.dartois@u-psud.fr}\\
%         \and
$^2$                Institut de Physique Nucl\'eaire d'Orsay (IPNO), CNRS-IN2P3, Universit\'e de Paris-Sud, Universit\'e Paris-Saclay, F-91406 Orsay, France\\
%          \and
$^3$               Centre de Recherche sur les Ions, les Mat\'eriaux et la Photonique, CIMAP-CIRIL-GANIL, Normandie Universit\'e, ENSICAEN, UNICAEN, CEA, CNRS, F-14000 Caen, France\\
%            \and
 $^4$              Centre de Sciences Nucl\'eaires et de Sciences de la Mati\`ere (CSNSM), CNRS/IN2P3, Universit\'e de Paris-Sud, Universit\'e Paris-Saclay, F-91405 Orsay, France\\
%            \and
 $^5$              Department of Physics, Indian Institute of Technology, Hauz Khas, New Delhi 110016, India.\\
 \vspace*{0.5cm}
{\it \large To appear in Astronomy \& Astrophysics}
 %            \thanks{The university of heaven temporarily does not accept e-mails}
 %            }
%
%   \date{Received April 22, 2018; accepted July 30, 2018}
%
% \abstract{}{}{}{}{} 
% 5 {} token are mandatory
 \vspace*{0.5cm} 
 }]

  \section*{Abstract}
  % context heading (optional)
  % {} leave it empty if necessary  
   {The occurrence of complex organic molecules (COMs) in the gas phase at low temperature in the dense phases of the interstellar medium suggests that a non-thermal desorption mechanism is at work because otherwise, COMs should condense within a short timescale onto dust grains. Vacuum ultraviolet (UV) photodesorption has been shown to be much less efficient for complex organic molecules, such as methanol, because mostly photoproducts are ejected. The induced photolysis competes with photodesorption for large COMs, which considerably lowers the efficiency to desorb intact molecules.}
  % aims heading (mandatory)
   {We pursue an experimental work that has already shown that water molecules, the dominant ice mantle species, can be efficiently sputtered by cosmic rays. We investigate the sputtering efficiency of complex organic molecules that are observed either in the ice mantles of interstellar dense clouds directly by infrared spectroscopy (CH$_3$OH), or that are observed in the gas phase by millimeter telescopes (CH$_3$COOCH$_3$) and that could be released from interstellar grain surfaces.}
  % methods heading (mandatory)
   {We irradiated ice films containing complex organic molecules (methanol and methyl acetate) and water with swift heavy ions in the electronic sputtering regime. We monitored the infrared spectra of the film as well as the species released to the gas phase with a mass spectrometer.}
  % results heading (mandatory)
   {We demonstrate that when methanol or methyl acetate is embedded in a water-ice mantle exposed to cosmic rays, a large portion is sputtered as an intact molecule, with a sputtering yield close to that of the main water-ice matrix. This must be even more true for the case of more volatile ice matrices, such as those that are embedded in carbon monoxide.}
  % conclusions heading (optional), leave it empty if necessary 
   {Cosmic rays penetrating deep into dense clouds provide an efficient mechanism to desorb complex organic molecules. Compared to the VUV photons, which are induced by the interaction of cosmic rays, a large portion desorb as intact molecules with a proportion corresponding to the time-dependent bulk composition of the ice mantle, the latter evolving with time as a function of fluence due to the radiolysis of the bulk.}
%136_Xe_23+_0_7_MeV_u

%   \keywords{Astrochemistry, cosmic rays, molecular processes, lines and bands,  interstellar ice mantles, solid state: volatile}

%   \maketitle
%
%________________________________________________________________

\section{Introduction}
The observed relatively high abundance of complex organic molecules in dense and pre-stellar phases of the interstellar medium is puzzling 
\citep[e.g.][and references therein]{Lefloch2018, Soma2018, Ceccarelli2017, Lopez-Sepulcre2017, Jorgensen2016, Jimenez-Serra2016, Oberg2014, Vasyunina2014, Bacmann2012}. If these molecules are formed on the surface of dust grains, an effective desorption mechanism must be invoked for their release in the gas phase.
As shown in laboratory experiments, many of the desorption mechanisms working effectively on small icy molecules such as CO and N$_2$ become much less relevant for large molecular systems comprising about   six or more atoms, the complex organic molecules (COM)s. In addition, the source of energy required for the desorption in the case of vacuum ultraviolet (VUV) photons easily photolyses large molecules such as methanol \citep[e.g.][]{Cruz-Diaz2016,Bertin2016}.
Thermal desorption of COMs is also much less efficient for large molecular systems because of the out-diffusion of the most volatile species without efficient transport of COMs \citep[e.g.][]{Ligterink2018}. Competition with other channels during formation \citep[e.g.][]{Minissale2016} is also of consequence
%\LEt{"important" means "of consequence"; do you mean "strong"? please check throughout and rephrase where necessary} 
at the surface of the grains in the desorption of the smaller species.

This article reports the experimental measurement of the desorption of COMs by sputtering following the impact of cosmic rays on ice mantles. 
The focus is on the most abundant well-known 6-atom ice mantle COM, methanol (CH$_3$OH). We also investigate an even larger COM with 11 atoms, methyl acetate (CH$_3$COOCH$_3$), which has been detected quite recently in Orion \citep[][]{Tercero2013}. The experiments are described in section 2. Section 3 presents the results of a combined IR and mass spectrometer analysis of the products in the experimentally simulated ice mantles that are ejected upon the swift interaction of heavy ions with thin ice films that are deposited at low temperature. Radiolysis and sputtering are examined together within a simple model framework. In the last section the observed astronomical occurrence and variations with respect to water ice and carbon monoxide ices of methanol in interstellar ice mantles are considered. Before we conclude, we discuss the destruction cross-section dependency on stopping-power.
%\LEt{please check that I got this right;\ is there a way to phrase this less intimidatingly? "we discuss the dependence of the stopping power on the destruction cross-section"?} 
and the astrophysical implications of sputtering rates for COMs.

%
%____________________________________________
\begin{figure}[tbhp]
\includegraphics[width=\linewidth]{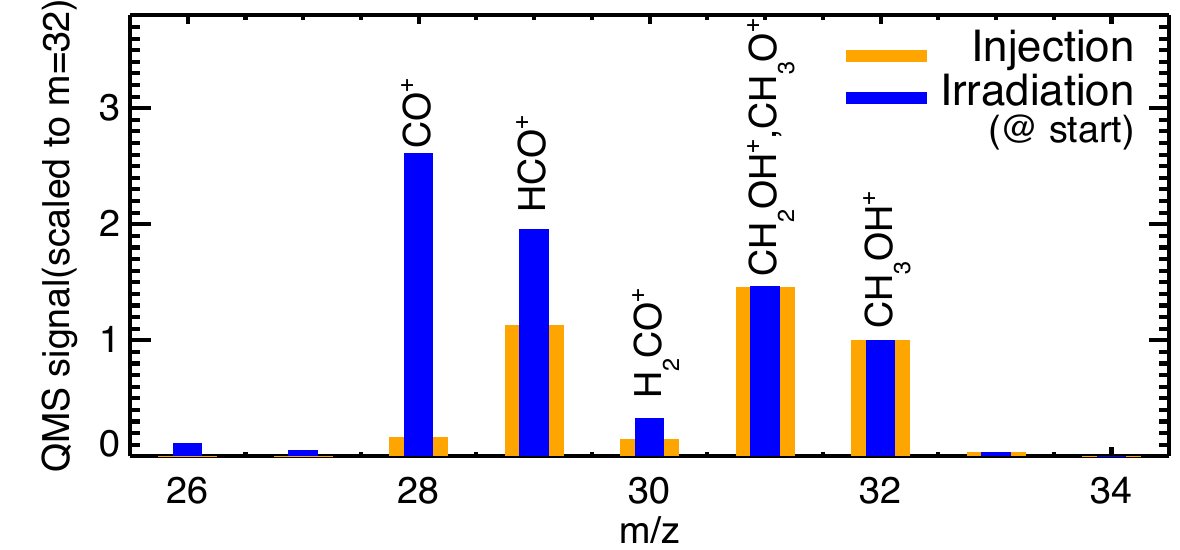}
\caption{Average of the mass spectra that were recorded during the methanol injection that was meant to form the ice film (orange histogram), tracing the QMS mass-fragmentation pattern of pure methanol; average of a few spectra that were recorded at the beginning of the ice-film irradiation (blue histogram). Both mass spectra are scaled to mass 32 for comparison. This shows that masses 32 and 31 can be used to monitor methanol because the radiolysis with the ion mainly contributes to peaks at lower masses. See text for details.} \label{fig:fragmentation_patterns_29W3}
\end{figure}
%____________________________________________
%
%_____________________________________________________________
%                                             Two column Table 
%_____________________________________________________________
%
\begin{table*}
\caption{Summary of experiment parameters}             
\label{table:1}                
\begin{center}
\begin{tabular}{l c c c c c l l }     % 8 columns 
\hline\hline       
                      % To combine 4 columns into a single one 
\#exp. tag      &T              &N$\rm _0(H_2O)$$^a$            &N$\rm _0(CH_3OH)$$^b$         & CH$_3$OH/H$_2$O$^c$   &QMS range      &Displayed      &Destr. cross section         \\%&Aspect\\%& \multicolumn{4}{c}{Method\#3}\\ 
                &K              &[10$^{16}$cm$^{-2}$]           &[10$^{16}$cm$^{-2}$]           &                                                 &amu            &in Fig.        &[10$^{-13}$cm$^{2}$]   \\
\hline                    
29W3    & 10            &-                                              &317.7                                  & -                                               &1-50           &\ref{fig:mosaic_29W3_28_W2_BIS_28_W3_BIS} (upper) &5$\pm$1        \\
28W2BIS  & 10           &-                                              &64.3                                   & -                                               &1-50           &\ref{fig:mosaic_29W3_28_W2_BIS_28_W3_BIS} (middle)        &5.9$\pm$1.5    \\
28W3BIS  & 10           &-                                              & 7.52                                    & -                                             &1-50           &\ref{fig:mosaic_29W3_28_W2_BIS_28_W3_BIS} (lower) &       \\
28W3    & 10            &20.6                                   &1.23                                   & 0.06                                    &1-33           &\ref{fig:mosaic_28W3}  &       \\
29W3BIS  & 10           &14.5                                   &17.8                                   & 1.22                                    &1-33           &\ref{fig:mosaic_29W3_BIS}      &       \\
%
%%%%%%%%%%%%%%%%%%%%%%%%%%%%%%%%%%%%%%%%%%%%%%%%%%%%%%%%%%%%%%%%%%%%
%
\hline\hline                    
\#exp. tag      &T              &N$\rm _0(H_2O)$$^a$            &N$\rm _0(CH_3OOCH_3)$$^b$         & CH$_3$OOCH$_3$/H$_2$O$^c$     &QMS range      &Displayed      &         \\%&Aspect\\%& \multicolumn{4}{c}{Method\#3}\\ 
                &K              &[10$^{16}$cm$^{-2}$]           &[10$^{16}$cm$^{-2}$]           &                                                 &amu            &in Fig.        &       \\
\hline                    
29W2BIS  & 10           &-                                              &6.15                                   & -                                               &1-80           &\ref{fig:mosaic_29W2_BIS}      &7$\pm$1.5      \\
29W1BIS  & 10           &15.1                                   &1.95                                   & 0.13                                    &1-80           &\ref{fig:mosaic_29W1_BIS}      &       \\
%
% thickness (cm) = N0 (cm^-2)  / [NA(molecule/mol)/18(g/mol)*0.9338(g/cm^3)]
% thickness (molecules) = N0 (cm^-2)  / [NA(molecule/mol)/18(g/mol)*0.9338(g/cm^3)]^(2./3)
%
\hline                  
\end{tabular}
\end{center}
Column density uncertainties are dominated by the uncertainties of integrated band strengths given in Table~\ref{table:2}.
$^a$  From the O-H stretching mode, with an adopted integrated absorption cross section of $\rm A=2.2\times 10^{-16}$ cm/molecule. 
When methanol becomes an important fraction of the mixture, the methanol contribution to the O-H is evaluated from its C-O stretching mode, and then subtracted, adopting an integrated absorption cross section of $\rm A(C-O)=1.5\times 10^{-17}$ cm/molecule and $\rm A(OH)=1.1\times 10^{-16}$ cm/molecule.
$^b$  From the C-O stretching mode, with an adopted integrated absorption cross section of $\rm A=1.5\times 10^{-17}$ cm/molecule. 
$^c$ initial fraction before irradiation starts. 
\end{table*}
%
%_____________________________________________________________
%                                             End Two column Table 
%_____________________________________________________________
%
%_____________________________________________________________
%                                             Two column Table 
%_____________________________________________________________
%
\begin{table*}
\caption{Integrated band strengths used in the analysis}             
\label{table:2}                
\begin{center}
\begin{tabular}{l l l l l }     % 8 columns 
\hline\hline       
Species                         &Mode                   &Position                               &A                              &Ref            \\
                                        &                               &cm$^{-1}$                      &cm.molec$^{-1}$        &               \\
\hline
CO                                      &CO stretch             &2140                   &$\rm 1.1\times10^{-17}$                      &\cite{Jiang1975}       \\
                                        &                               &                                 &$\rm 1.1\times10^{-17}$                        &\cite{Gerakines1995}   \\
                                        &                               &                               &$\rm 1.12\times10^{-17}$             &\cite{Bouilloud2015}   \\
                                        &                               &                                 &$\rm 1.1\times10^{-17}$                        &{\bf\it Adopted for this work}          \\
CO$_2$                          &CO$_2$ stretch &2350                   &$\rm 7.6\times10^{-17}$                      &\cite{Gerakines1995} \\
H$_2$O                          &OH stretch             &3600-2700              &$\rm 2.2\times10^{-16}$                      &\cite{ldh1986} \\
                                        &                               &                                 &$\rm 2.2\pm^{0.}_{0.2}\times10^{-16}$  &{\bf\it Adopted for this work}          \\
CH$_3$OH                        &OH stretch             &3600-2700              &$\rm 1.1\times10^{-16}$                      &\cite{ldh1986} \\
                                        &                               &                                 &$\rm 1.28\times10^{-16}$               &\cite{Palumbo1999}             \\
                                        &                               &                                 &$\rm 1.0\times10^{-16}$                        &\cite{Bouilloud2015}           \\
                                        &                               &                                 &$\rm 1.1\pm0.15\times10^{-16}$ &{\bf\it Adopted for this work}          \\
CH$_3$OH                        &C-O stretch            &1030                   &$\rm 1.8\times10^{-17}$                      &\cite{ldh1986} \\
                                        &                               &                               &$\rm 1.8\times10^{-17}$                      &\cite{Sandford1993}    \\
                                        &                               &                               &$\rm 1.2\times10^{-17}$                      &\cite{Palumbo1999}             \\
                                        &                               &                               &$\rm 1.07\times10^{-17}$             &\cite{Bouilloud2015}           \\
                                        &                               &                                 &$\rm 1.5\pm^{0.3}_{0.4}\times10^{-17}$ &{\bf\it Adopted for this work}          \\
CH$_3$COOCH$_3$ &C-O stretch            &1255                   &$\rm 5\times10^{-17}$~$^a$             &\cite{ldh1986} \\
                                        &                               &                                 &$\rm 5.\pm0.\times10^{-17}$            &{\bf\it Adopted for this work}          \\
\hline                  
\end{tabular}
\end{center}
$^a$ band strength from the CO stretching mode of ethyl acetate measurement
%$^a$ liquid measurement
\end{table*}
%
%_____________________________________________________________
%                                             End Two column Table 
%_____________________________________________________________
%
%
\section{Experiments}
Swift ion-irradiation experiments were performed at the heavy-ion accelerator Grand Acc\'el\'erateur National d'Ions Lourds (GANIL, Caen, France).
Heavy-ion projectiles were delivered on the IRRSUD beam line\footnote{http://pro.ganil-spiral2.eu/laboratory/experimental-areas}. The Irradiation de GLaces d'Int\'er\^et AStrophysique (IGLIAS) facility, a vacuum chamber (10$^{-9}$ mbar under our experimental conditions) that holds an IR-transmitting substrate that can be cryocooled down to about 10 K, was coupled to the beam line. The ice films were produced by placing the cold window substrate in front of a dosing needle that was connected to the deposition line. Ice films were condensed at 10~K on the window from the vapour phase and were kept at this temperature during the irradiations.
Methanol was rectapur grade, and
methyl acetate (99.5\% purity) was purchased from Sigma Aldrich; both were used as received.
Details of the experimental setup are given in \cite{Auge2018}. 
The stopping power of  the projectiles ($\rm ^{136}Xe^{23+}$ at 0.7 MeV/u) in the electronic regime is close to 8keV/nm for a pure H$_2$O ice film when we adopt an ice density of 0.93 g/cm$^3$.
For pure methanol, the stopping power is slightly higher, about 9.4keV/nm for a density of pure ice of 1.013g/cm$^3$ \citep[e.g.][]{Mate2009}, and is about 7.5 to 8.3keV/nm for methyl acetate when a pure ice density of between 0.9 to 1.0g/cm$^3$
is assumed. This stopping-power value lies in a range close to previous thick-film experiments \citep[][and references therein]{Dartois2015, Dartois2018}. 
%\LEt{please check your LaTeX command for the reference; the "and references therein" should be within the parentheses. This is a LaTeX command error that I cannot fix for you with the program I work with}. 
%The same projectiles as were used in the study
%\LEt{"this study" here means this paper that I'm currently editing. If you wish to refer to your previous studies, you need to repeat them here (and make clear whether it's the 2015 or the 2018 paper) because you cannot refer like this to information that was given in parentheses} 
%
%were employed to 
The determination of the semi-infinite sputtering yield was discussed in \cite{Dartois2018}, using the same projectiles. This previously determined yield-dependency on the stopping power for pure water ice predicts a total sputtering yield for intact plus radiolysed molecules
%{\bf (including intact and radiolysed molecules, but largely dominated by intact molecules\LEt{it is not clear what the second part refers to, please rephrase. Are the intact and radiolysed molecues largeley dominated...? then it should be "... radiolysed molecules, which are largely..." or do you mean the sputtering yield? then it should be "... radiolysed molecues; the yield is largely dominated..."})}
 of 2.0$\pm$0.2$\times$10$^{4}$ sputtered H$_2$O/ion; this yield is dominated by intact molecules.
The ion flux, set between 10$^7$ to 10$^9$ ions/cm$^2$/s, was monitored on-line using the current measured on the beam entrance-slits that define the aperture. 
The relation between the current at different slit apertures and the flux was calibrated before the experiments. We used a Faraday cup that was inserted in front of the sample chamber to do this. 
The deposited ice-film thicknesses allow the ion beam to pass through the film with an almost constant energy loss per unit path length.
A Bruker FTIR spectrometer (Vertex 70v) with a spectral resolution of 1 cm$^{-1}$ was used to monitor the IR film transmittance. The evolution of the IR spectra was recorded as a function of the ion fluence.
The irradiation was performed at normal incidence, whereas the IR transmittance spectra were recorded simultaneously at 12$\rm^o$ of incidence (a correction factor of $\approx$0.978 was therefore applied to determine the normal column densities).
A sweeping device allows for uniform and homogeneous ion irradiation over the target surface.
Mass measurements were performed simultaneously using a microvision2 mks quadrupole mass-spectrometer (QMS). The QMS  signals were noise-filtered to smooth out high-frequency temporal fluctuations using a Lee filter algorithm with a typical box size of five to seven points.
The mass ranges that we scanned were varied for the different experiments in order to optimise the integration time.
The masses we used to follow a given species were selected in order to avoid overlap with masses that are dominated by contamination and avoid confusion with a byproduct of the radiolysis.
For water, masses 17 and 18 were used.
For methanol, mass 31 was used preferentially, especially when the signal was low. It is the strongest signal in the high-mass fragmentation pattern for methanol and was recorded for pure methanol during injection, as shown in Fig.\ref{fig:fragmentation_patterns_29W3}. Mass 32 can also be used, but it will be contaminated by radiolytically produced oxygen in the mixtures that contain high proportions of H$_2$O. Masses 28, 29, and 30 contain significant fractions of CO and H$_2$CO fragments, which are the main products of methanol radiolysis at the surface and/or in the bulk of the ice \citep[see the summary Table 6 in][]{Barros2011}. Mass 31 can contain small amounts of methanol radiolysis products such as metoxy or hydroxymethyl radicals (CH$_3$O, CH$_2$OH), but as shown in Fig.\ref{fig:fragmentation_patterns_29W3} for the first steps of irradiation and also in the methanol experiments figures, the m=32/m=31 ratio
%\LEt{is this "signal-to-noise ratio"? if so, please use this (or the abbreviated form, "S/N", after introducing this as an abbreviation. If you choose to use the abbreviation, please make sure that you use it consistently throughout, except for the beginnings of sentences, where there should be no abbreviation}
is almost constant and close to the pure methanol value that is recorded systematically with the QMS before irradiation. In these experiments, the direct contributions of the metoxy or hydroxymethyl radical should therefore be considered as low.
When only selected optimum masses are used, a correction factor needs to be applied to estimate the abundance of a species so that the mass fragmentation pattern following electron impact ionisation can be taken into account. The experimental fragmentation pattern within our QMS for a species X was monitored during the injection of the gas mixture, that is, when the ice film was deposited. A self-calibration of the QMS for the mass-fragmentation pattern of species X was then obtained at mass m as follows:
\begin{equation}
\rm f(X,m)=I(m)/\sum_{m=Xfragments}I(m)
,\end{equation}
where only the expected masses from possible (major) fragments (X$\rm _{fragments}$) were included, for instance, masses 16, 17, and 18 for H$_2$O.
Therefore, the sum of the chosen masses were used and divided by their relative fragmentation-pattern percentage to retrieve the total number of species X.
The spectra were also corrected for the total electron-impact ionisation cross section $\rm \sigma^{impact}(X)$ at 70 eV (energy of the QMS electron ionisation source) for each molecule: H$_2$O (2.275~$\AA^2$, NIST database), CH$_3$OH \citep[4.8~$\AA^2$,][]{Vinodkumar2011}, and CH$_3$OOCH$_3$ \citep[11.2~$\AA^2$,][] {Kaur2015}, to compensate for the higher ionisation efficiency of larger species (which carry more electrons).

The abundance ratios of species X and Y were thus evaluated with
\begin{equation}
\rm \frac{[X]}{[Y]} = \frac{\sum_{m=Xfragments}I(m)}{\sum_{m=Xfragments}f(X,m)}\times \frac{\sum_{m=Yfragments}f(Y,m)}{\sum_{m=Yfragments}I(m)} \times \frac{\sigma^{impact}(Y)}{\sigma^{impact}(X)}
,\end{equation}
with the fragments chosen so that they have a significant signal-to-noise ratio and do not overlap with other {dominant} species and/or potential residual gas masses. 
%
%%%%%%%%%%%%%%%%%%%%%%%%%%%%%%%%%%%%%%%%%%%%%%%%%%%%%%%%%%%
%
% BIG BIG FIGURE
%____________________________________________
\begin{figure*}[tbhp]
\begin{minipage}[c]{1.1\columnwidth}
\includegraphics[width=\linewidth]{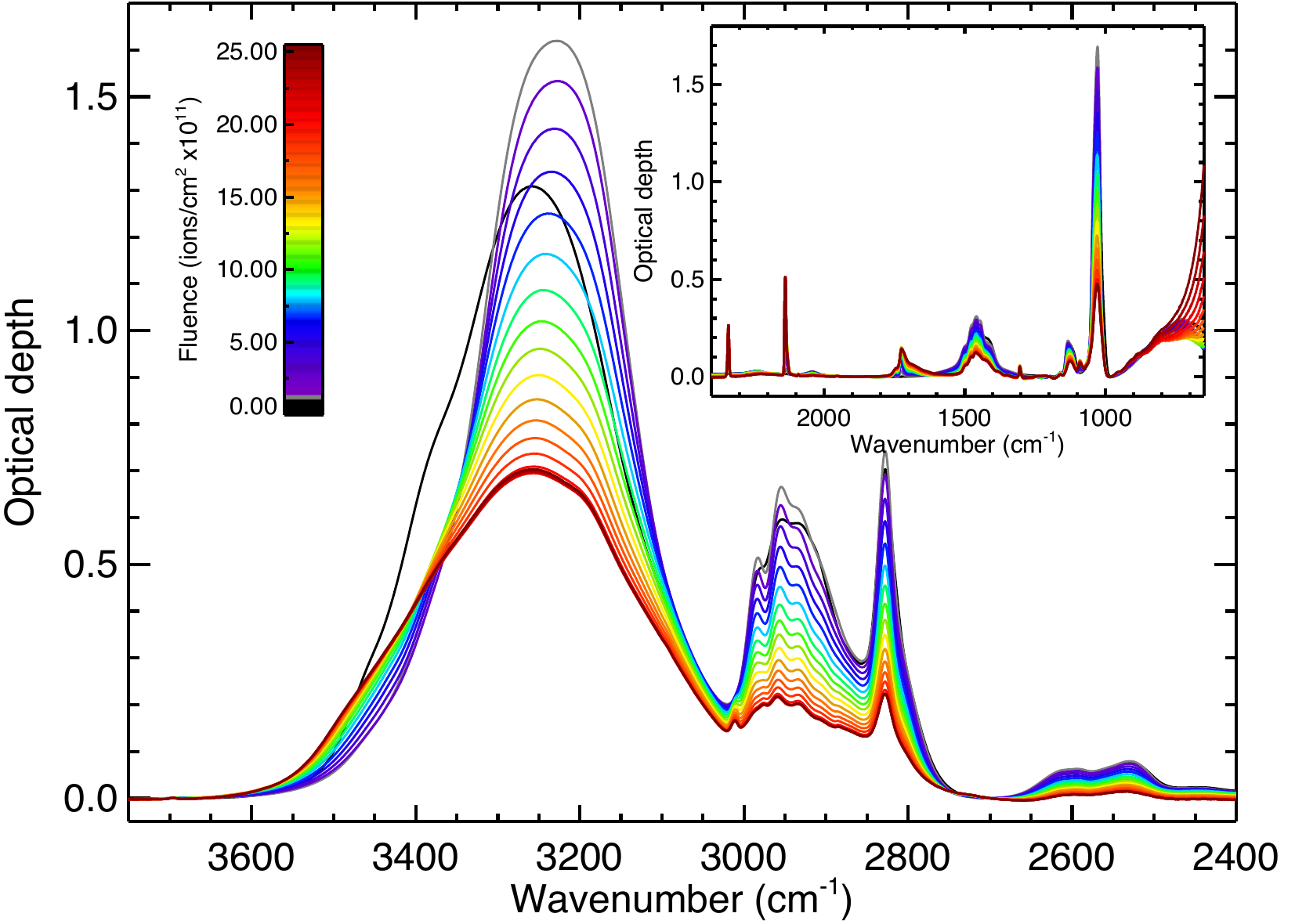}
\end{minipage}
\begin{minipage}[c]{0.89\columnwidth}
\includegraphics[width=\linewidth]{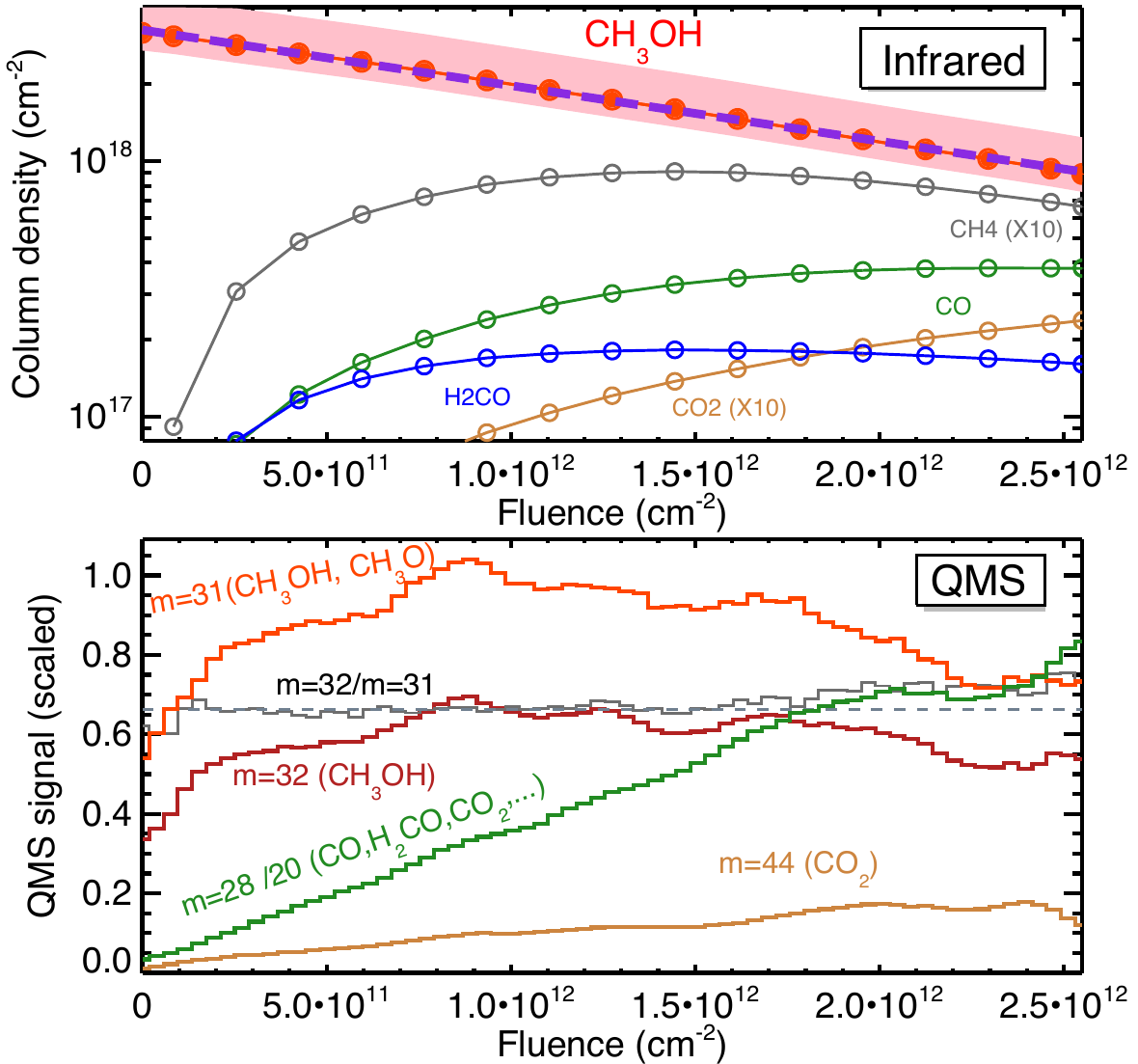}
\end{minipage}
\begin{minipage}[c]{1.1\columnwidth}
\includegraphics[width=\linewidth]{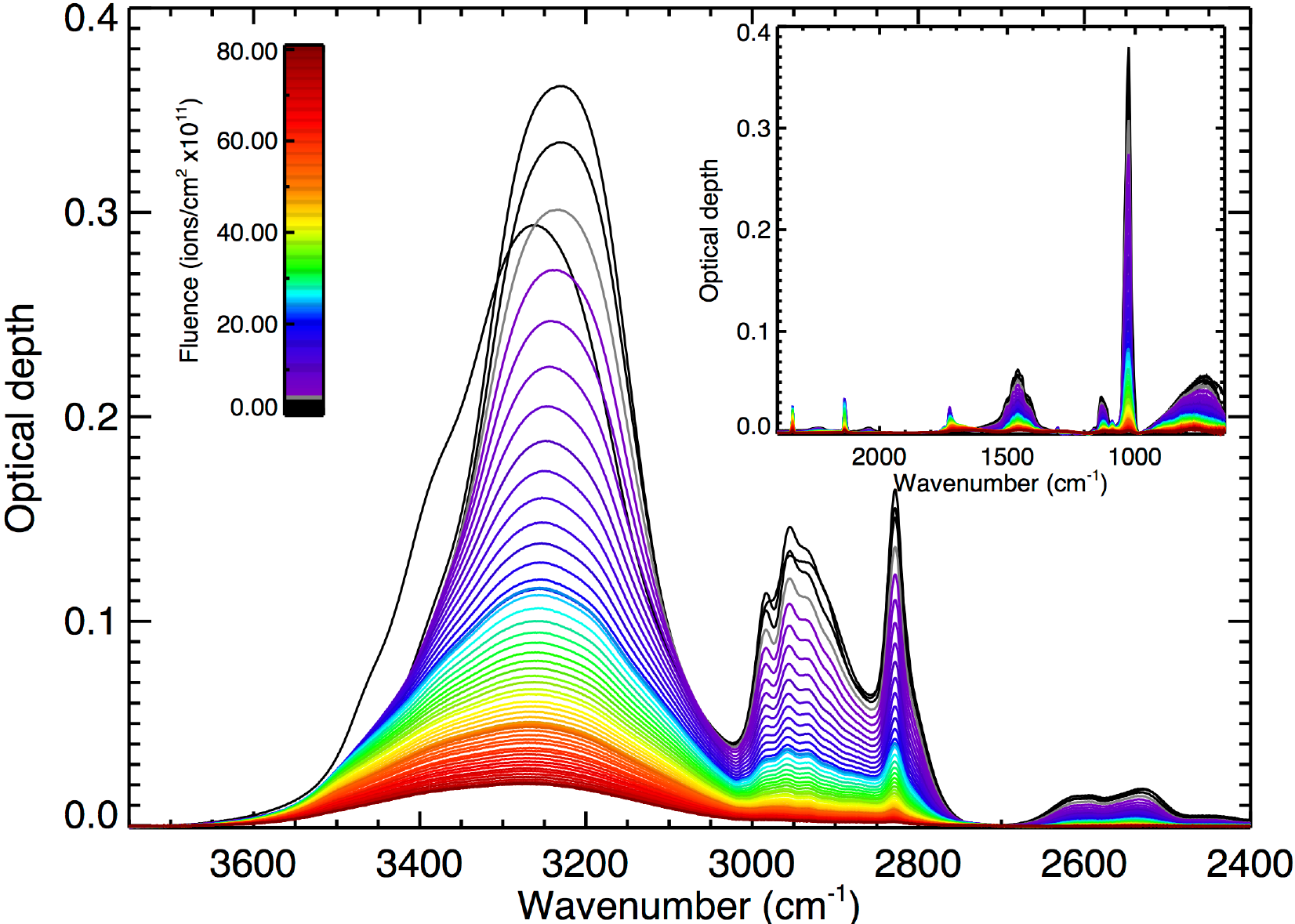}
\end{minipage}
\begin{minipage}[c]{0.89\columnwidth}
\includegraphics[width=\linewidth]{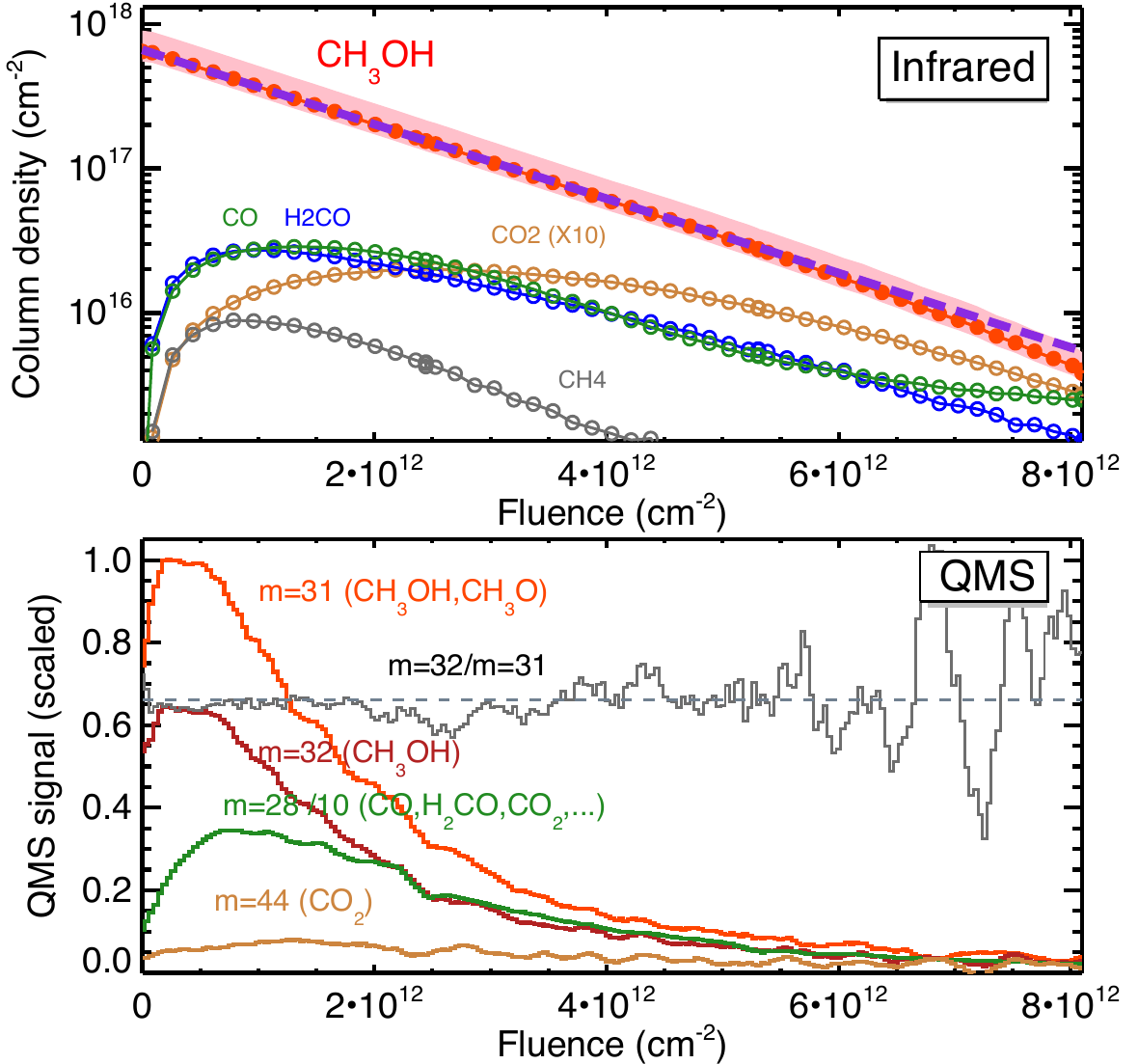}
\end{minipage}
\begin{minipage}[c]{1.1\columnwidth}
\includegraphics[width=\linewidth]{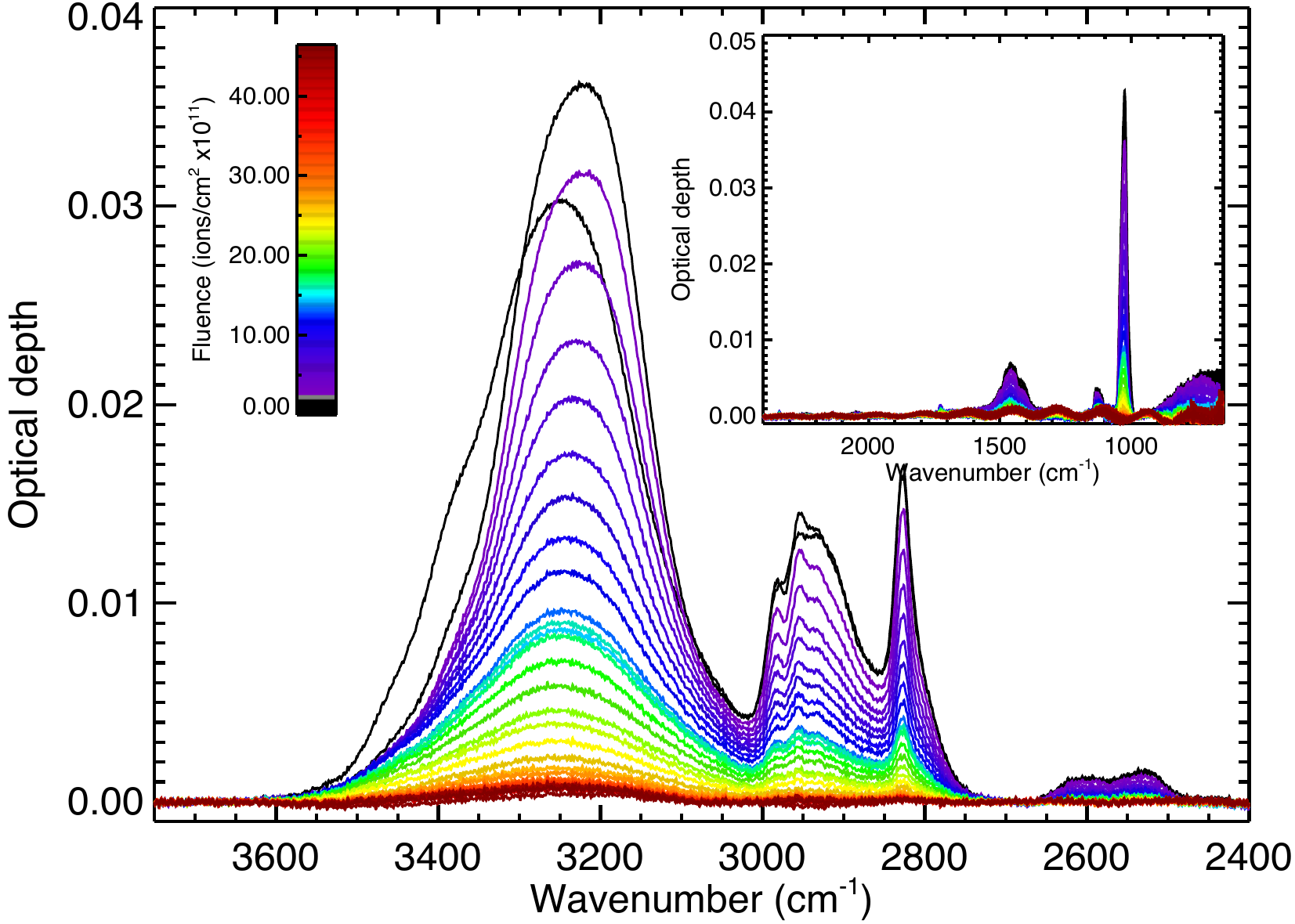}
\end{minipage}
\begin{minipage}[c]{0.04\columnwidth}
$ $
\end{minipage}
\begin{minipage}[c]{0.89\columnwidth}
\includegraphics[width=\linewidth]{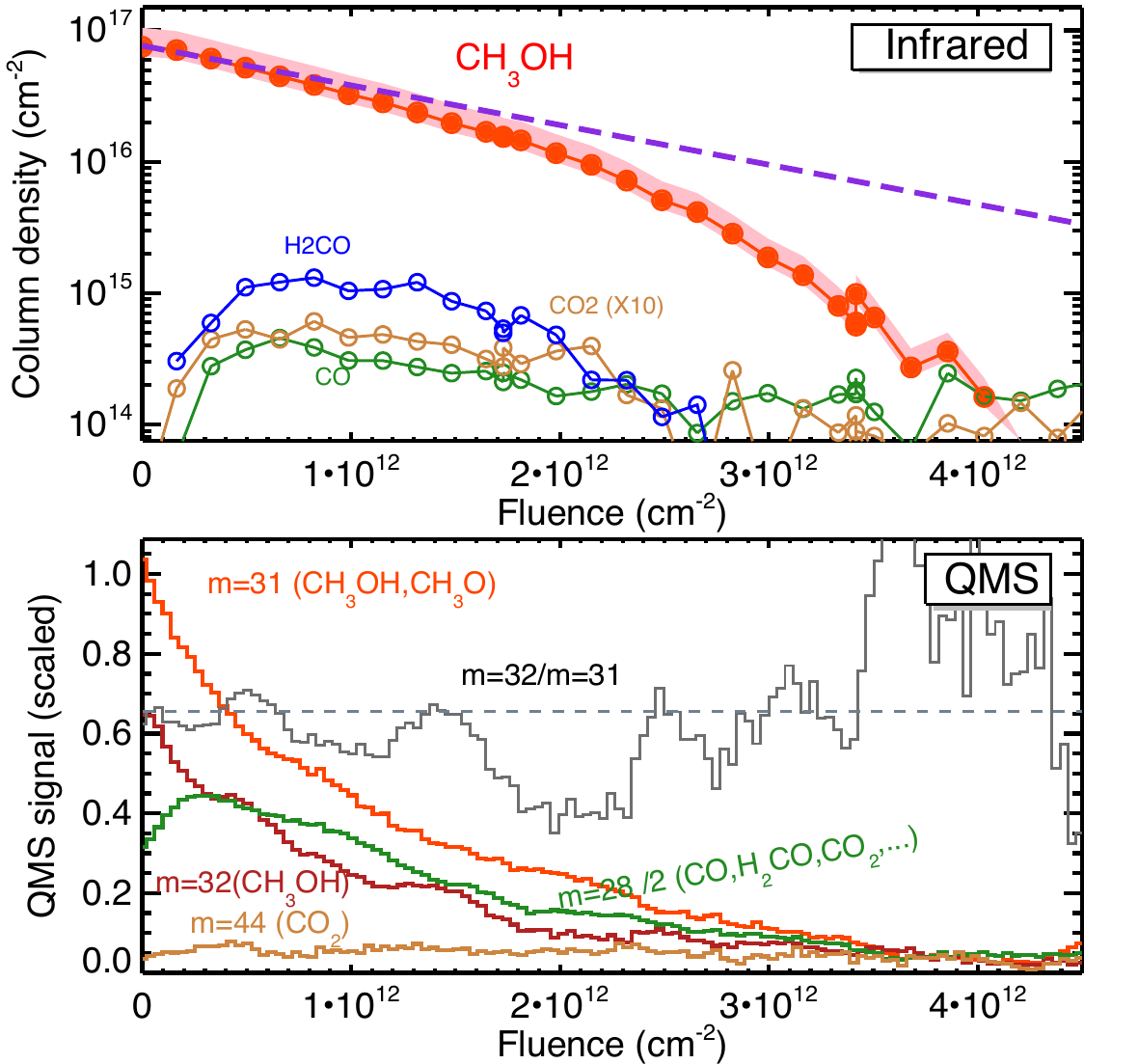}
\end{minipage}
\caption{Pure methanol ice-film experiments with various initial film thicknesses (thick to thin from top to bottom). Left panels: Evolution of the CH$_3$OH IR spectra as a function of fluence.
Right top panels: Ice column density measurements from IR spectra (circles). The column densities are estimated using the integrated cross sections from Table~\ref{table:2}. The purple dashed line is the fit to the destruction cross section of methanol (top $\rm 5\pm1\times10^{-13}~cm^2/ion$, middle $\rm 5.9\pm1\times10^{-13}~cm^2/ion$).
For the thinner film (bottom panels), the onset of dominant sputtering appears for a column density lower than $\approx$~$10^{16}$~cm$^{-2}$.
Right lower panels: QMS scaled intensities at selected masses that were used to follow carbon dioxide, methanol m=32, m=31, and m=28 (CO, H$_2$CO, and CO$_2$) during irradiation. {The mass-pattern ratio for methanol m=32/m=31 (grey curves) is compared to the ratio measured for pure methanol during the ice-film preparation (dotted line).} See text for details.
\label{fig:mosaic_29W3_28_W2_BIS_28_W3_BIS}}
\end{figure*}
%____________________________________________
%
%
%%%%%%%%%%%%%%%%%%%%%%%%%%%%%%%%%%%%%%%%%%%%%%%%%%%%%%%%%%%
%

For each of the pure methanol and methyl acetate experiments, the measured QMS mass signals are hereafter displayed multiplied by an arbitrary global factor (i.e. scaled to span a unit range). In the mixture experiments, in order to establish the relative abundance of species, the QMS signals are presented corrected for the relative mass-fragmentation pattern for a given mass or group of masses that is attributed to a dominant species, normalised by its electron-impact ionisation cross section, as discussed above. The signals are then scaled by a global factor.
During the experiments, the ion beam was regularly stopped during a few cycles in order to record and follow the evolution of the QMS and chamber background signal.
The QMS spectra we present are background subtracted.
A summary of the ice-film parameters such as ice mixture, deposition, and irradiation temperatures, and the scanned QMS masses is given in Table~\ref{table:1}.
A summary of the integrated band strengths from the literature that we used in the IR spectra analysis to determine ice-film column densities is listed in Table~\ref{table:2}.
%
%%%%%%%%%%%%%%%%%%%%%%%%%%%%%%%%%%%%%%%%%%%%%%%%%%%%%%%%%%%%%%%%%%%%%
%
%____________________________________________
\begin{figure*}[tbhp]
\begin{minipage}[c]{1.1\columnwidth}
\includegraphics[width=\linewidth]{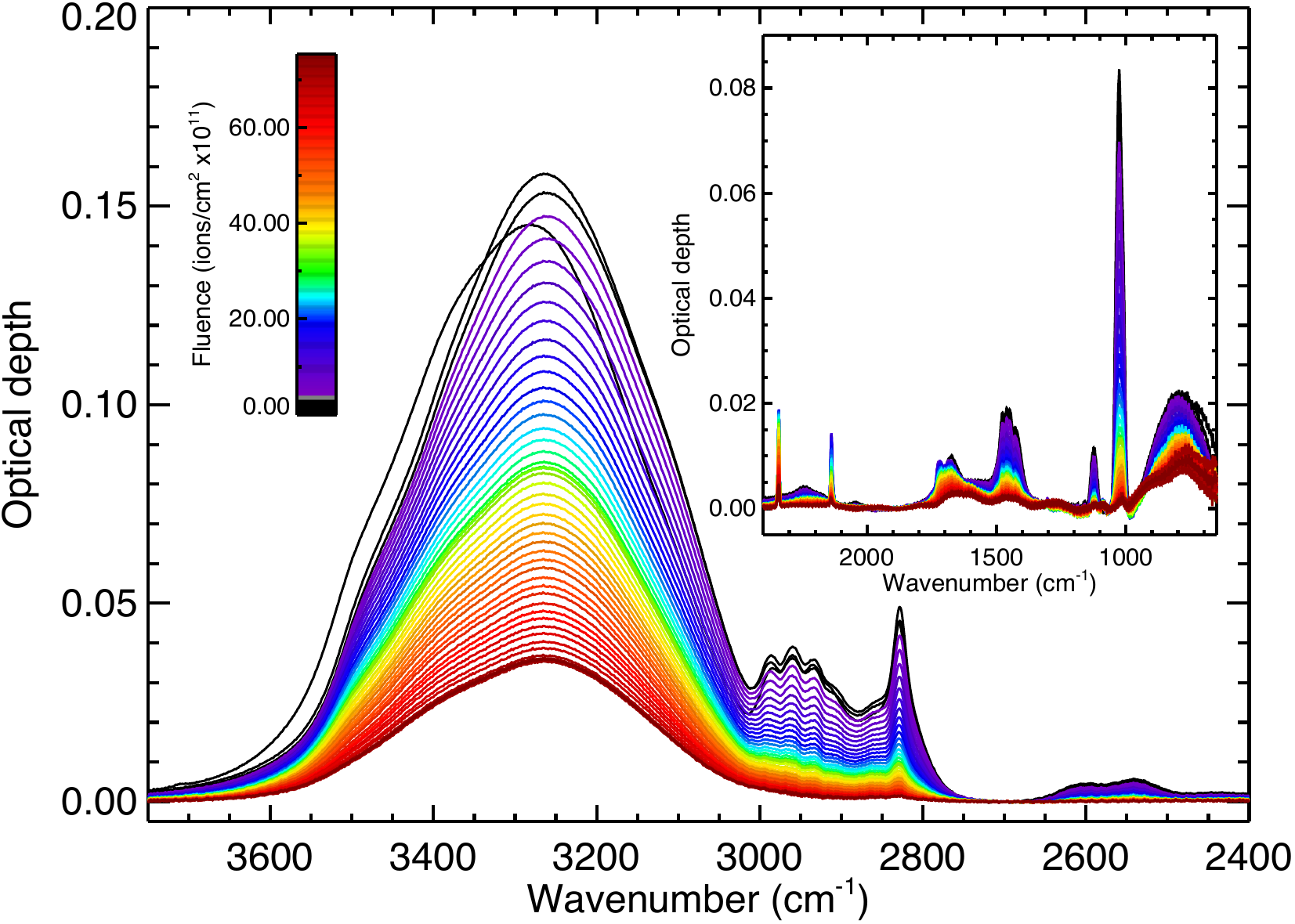}
\caption{Left: H$_2$O:CH$_3$OH (100:122) ice film. Evolution of the IR spectra as a function of fluence.
Right top panel: Ice column density measurements from IR spectra. The black dashed lines represent the fits to the destruction cross section of water and methanol.
Middle panel: QMS-normalised signals we used to follow the relative abundance of water (m=17+m=18), methanol (m=32 and m=31), and carbon monoxide (m=28) upon irradiation. See text for details.
m=28 increases at the beginning because it is mainly composed of fragments from the radiolytic products of the initial ice mixture.
Lower panel: CH$_3$OH/H$_2$O abundance ratio deduced from the IR spectra of the ice film as a function of fluence (red dots, the uncertainty is filled in in red). Comparison with the QMS-determined abundance ratio of the same desorbed molecules (dark blue line, from m=32+m=31 for methanol and m=17+m=18 for water). The blue and cyan
%\LEt{the slash indicates "ratio", as in "S/N"\ for "signal-to-noise ratio". It is ambiguous if you use it like you do here. If you do not mean "and" but "or", please change, of course} 
star indicates the gas ratio measurement with the QMS during the ice film deposition. The dashed blue and cyan line represents equation \ref{equation_QMS_29W3_BIS}; the uncertainty is filled in blue. \label{fig:mosaic_29W3_BIS}}
\end{minipage}
\begin{minipage}[c]{0.05\columnwidth}
\end{minipage}
\begin{minipage}[c]{0.85\columnwidth}
\includegraphics[width=\linewidth]{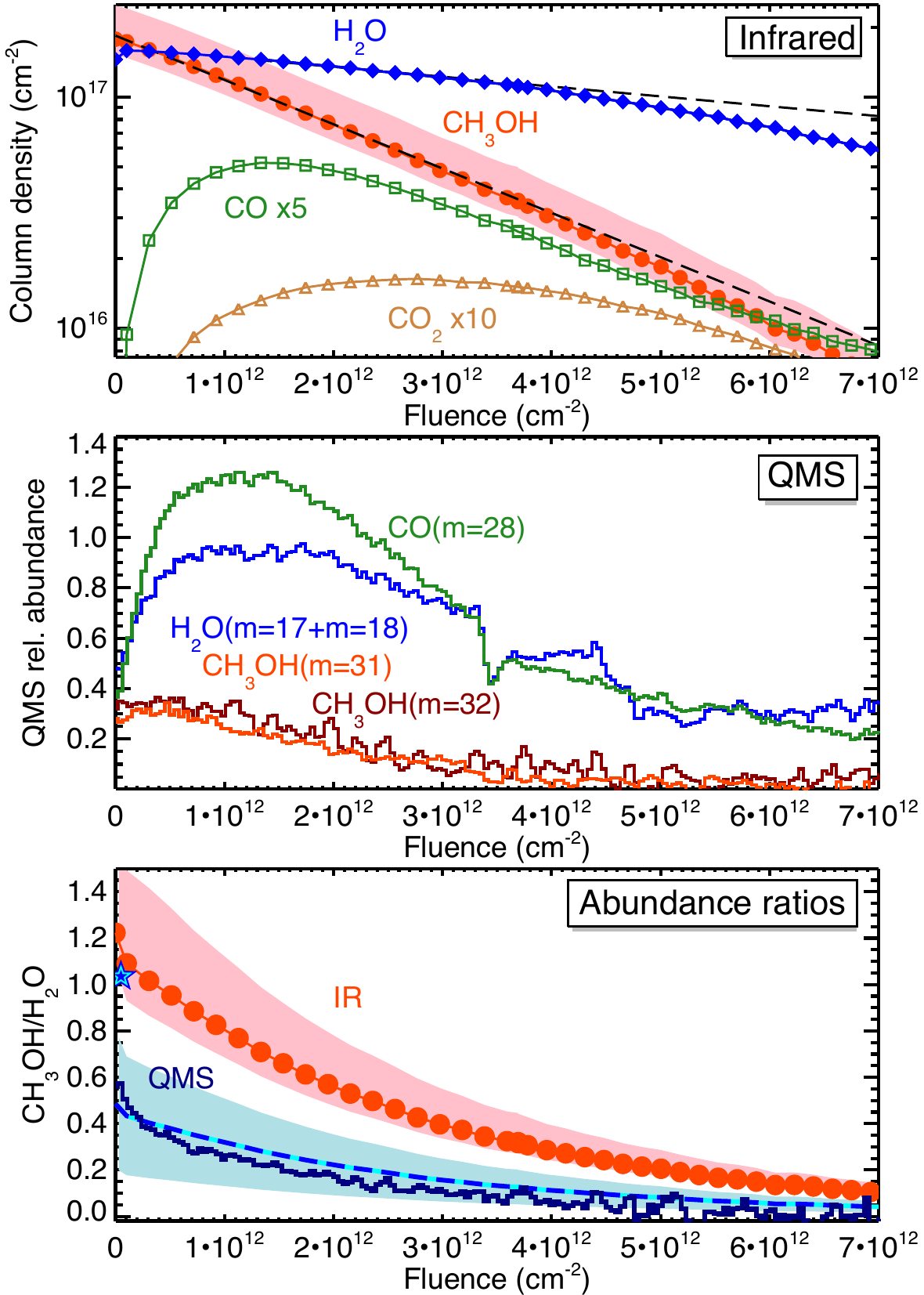}
\end{minipage}
\end{figure*}
%____________________________________________
%
%____________________________________________
\begin{figure*}[tbhp]
\begin{minipage}[c]{1.1\columnwidth}
\includegraphics[width=\linewidth]{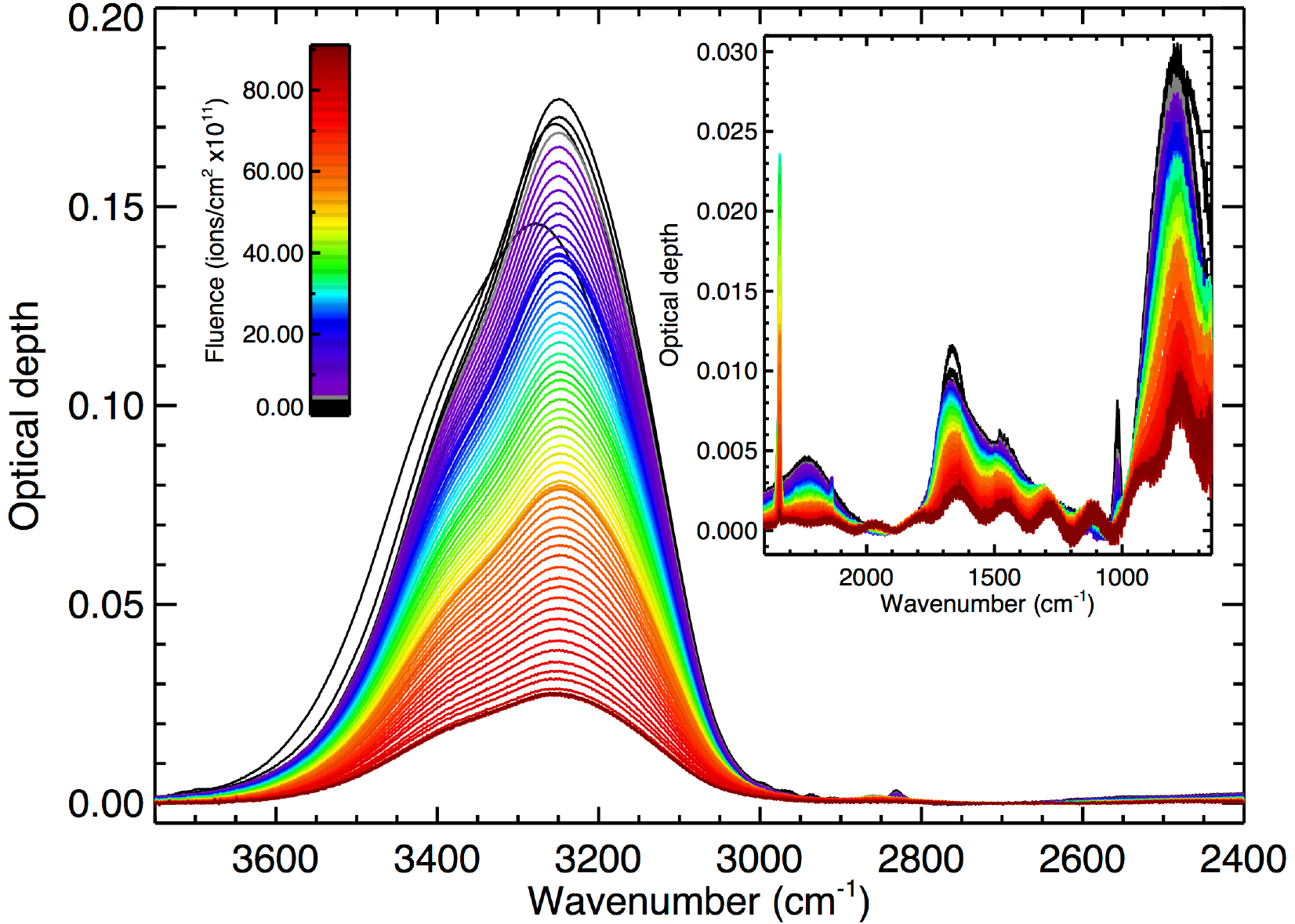}
\caption{Left: H$_2$O:CH$_3$OH (100:6) ice-film evolution of the IR spectra as a function of fluence.
Right: H$_2$O:CH$_3$OH 100:6 ice-film experiment. Top panel: Ice column density measurements from IR spectra. The black dashed lines represent the fits to the destruction cross section of water and methanol.
Middle panel: QMS-normalised signals we used to follow the relative abundance of water (m=17+m=18), methanol (m=31), and carbon monoxide (m=28) upon irradiation. See text for details.
 CO increases initially because it is a radiolytic product of the ice mixture.
Lower panel: CH$_3$OH/H$_2$O abundance ratio deduced from the IR spectra of the ice film as a function of fluence (red dots). Comparison with the QMS-determined abundance ratio of the same desorbed molecules (dark blue line). The blue and cyan star indicates the gas ratio measurement with the QMS during the ice-film deposition. The dashed blue and cyan line represents equation \ref{equation_QMS_28W3}; the uncertainty is filled in blue. \label{fig:mosaic_28W3}}
\end{minipage}
\begin{minipage}[c]{0.05\columnwidth}
\end{minipage}
\begin{minipage}[c]{0.85\columnwidth}
\includegraphics[width=\linewidth]{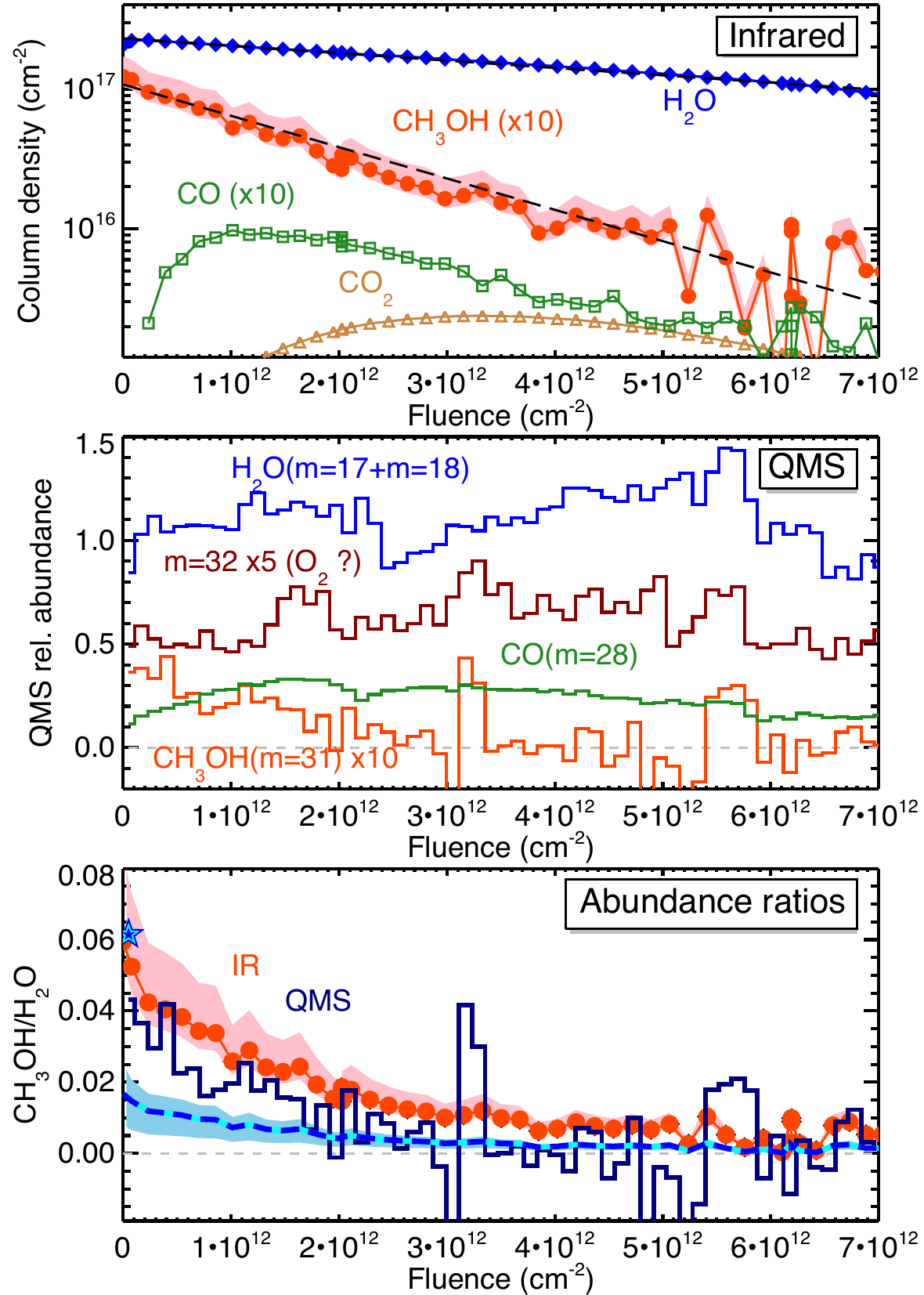}
\end{minipage}
\end{figure*}
%____________________________________________
%
%%%%%%%%%%%%%%%%%%%%%%%%%%%%%%%%%%%%%%%%%%%%%%%%%%%%%%%%%%%%%%%%%%%%%
%
\section{Results}
\label{results}
\subsection{Pure CH$_3$OH, thin and thick films}
Before we investigated potentially more relevant astrophysical methanol/ice abundance mixtures, we performed pure methanol ice-irradiation experiments for which we varied the film thicknesses to determine among other parameters the methanol destruction cross section under fixed ion-beam conditions. Three ice films with different thicknesses were irradiated. The evolution of the CH$_3$OH IR spectra, ice column densities, and QMS intensities as a function of fluence is displayed in Fig.\ref{fig:mosaic_29W3_28_W2_BIS_28_W3_BIS}.
The sputtering is unambiguously detected in the three experiments, and its contribution is negligible to the IR signal evolution compared to the radiolysed molecules for the two thicker ice experiments. This allows a direct determination of the radiolysis cross section.
For the two thicker ice films, a fit of the destruction cross section of methanol for the considered ion leads to two independent determinations of $\rm \sigma^{destruction}_{CH_3OH}=5\pm1\times10^{-13}~cm^2/ion$ and $\rm 5.9\pm1\times10^{-13}~cm^2/ion$.
The deviation from the apparent pure destruction evolution measured for thick films, that is, when the number of molecules that is removed by sputtering becomes comparable or higher in importance than the bulk destruction of methanol, is best observed when the film becomes thinner, here for column densities below a few $10^{16}$~cm$^{-2}$ , as shown in the lower panel of Fig.\ref{fig:mosaic_29W3_28_W2_BIS_28_W3_BIS}. 
The methanol sputtering yield can be calculated with the thinner ice experiment using a model similar to those used in
%\LEt{again, please check your LaTeX commands} 
\cite{Dartois2015, Dartois2018}. An estimate is obtained by adjusting a fit that includes the electron sputtering for the thinner film and setting the previously determined destruction cross section to $\rm 5.45\pm1.5\times10^{-13}~cm^2/ion$. We obtain $\rm Y_{CH_3OH}^{sputtering}\approx 1.4\pm0.6\times10^{4}$ sputtered CH$_3$OH/ion.
The QMS intensities are shown, scaled by a global factor, at selected masses that we used to follow carbon dioxide m=44, methanol m=32, m=31, and m=28 (mainly CO with lower contributions from H$_2$CO, CO$_2$) upon irradiation. The ratio of the m=32 to m=31 masses is shown as a function of the fluence. This ratio (grey) is almost constant and is compared to the mass-pattern ratio of pure methanol during the preparation of the ice film by injection of methanol in the chamber (dotted line) for each experiment. This constancy tells us that the contribution of the direct desorption of metoxy and/or hydroxymethyl radical species (CH$_3$O, CH$_2$OH) from the ice is at most a small contribution to the signal because it would imply a marked decrease in this ratio.
The relative intensities of the CO to CH$_3$OH masses from the QMS at a given fluence change with the initial ice thickness because CO is produced by the radiolysis and is also more mobile. Measured gaseous CO can arise from a greater depth than CH$_3$OH, and the CO to CH$_3$OH gaseous ratio at comparable fluences therefore decreases with the initial thickness of the film. This can also be inferred from the IR spectra because in the case of the thinner film, the radiolytically produced CO does not accumulate in the film, but desorbes almost entirely immediately from the film.
\subsection{CH$_3$OH:H$_2$O}
\subsection*{H$_2$O:CH$_3$O (100:122)}
A mixture of methanol and water-ice with equivalent proportions  (H$_2$O:CH$_3$OH-100:122) was irradiated. The recorded IR spectra and their evolution as a function of the ion fluence are shown in Fig.~\ref{fig:mosaic_29W3_BIS}. 
Using the C-O stretching mode of methanol and the OH stretch of water ice (corrected for the methanol OH contribution), we report the column density of these molecules in the ice.
The abundance ratio of gaseous methanol to water measured with the QMS during the ice deposition (blue star in Fig.\ref{fig:mosaic_29W3_BIS}, lower right panel) gives, within uncertainties, the same value as the ratio measured after deposition for the ice film in the IR (thus providing an independent measurement of the expected deposited ice-mixture ratio).
The abundance ratios of gas- and solid-phase methanol to water evolve in parallel during irradiation. The measured abundance ratio is lower than the bulk of the ice film, showing that the methanol sputtering yield is influenced both by its sputtering and by radiolysis efficiencies.
The mass spectra in Fig.\ref{fig:mosaic_29W3_BIS} show that water and methanol are desorbed as soon as the ion irradiation begins (middle panel) and only decrease, whereas the m=28 (gas-phase CO, CO$_2$, and H$_2$CO) intensity is lower at the beginning of irradiation, then rises, and finally decreases.
The observed gaseous CO arises partly as a direct sputtered CH$_3$OH radiolytic product and partly from the bulk of the ice at later times.
In addition, because it is much more highly volatile, the desorbing CO can come from deeper layers of the ice film than water and methanol. This is shown by the amount of m=28 that desorbs with respect to methanol, whereas the mean bulk ratio of these ices is lower (upper panel of Fig.\ref{fig:mosaic_29W3_BIS}) than what the QMS sees.
The extent of the radiolysis of methanol with respect to the sputtering of intact molecules 
is discussed in the next section.
\subsection*{Water-rich methanol ice, H$_2$O:CH$_3$OH (100:6)}
An experiment with an ice methanol water mixture was performed on a water-rich mixture, an H$_2$O:CH$_3$OH~(100:6) ice film. This mixture, which is dominated by water ice, was chosen for two reasons: (i) to represent a typical integrated abundance ratio of methanol to water ice, as is observed along many lines of sights, when methanol is detected. (ii) We expect the sputtering behaviour to be dominated by the main matrix component, water.
For this, the previous pure water-ice sputtering experiments can provide a good anchor point from which to start \citep[][]{Dartois2018}.
The recorded IR spectra and their evolution as a function of the ion fluence are shown in Fig.~\ref{fig:mosaic_28W3}. 
Selected QMS mass intensities are also shown, which allows us to monitor the gas-phase evolutions. For water, m=17 and m=18 were used. For methanol, only the m=31 mass was used. In contrast to the previous experiment, m=32 cannot be used because it is higher than expected from methanol fragmentation and its trend is flatter, closer to the trend of gas-phase water. This suggests that the m=32 mass is dominated by an O$_2$ (or H$_2$O$_2$ fragments) contribution that arises from the radiolysis of the main component of the ice, water. The other masses of the methanol fragmentation pattern are affected by other species (m=30, 29: H$_2$CO and fragments, m=28: CO and H$_2$CO fragments).
Mass m=28 was also followed to monitor the contribution of CO (main contributor, it also contains less important contributions from the H$_2$CO and CH$_3$OH fragmentation pattern). 
The abundance ratio of gaseous methanol to water measured with the QMS during the ice deposition (blue star in Fig.\ref{fig:mosaic_28W3}, lower right panel) gives, within uncertainties, a slightly higher value than the ratio measured after deposition for the ice film in the IR.
The abundance ratios of gas- and solid-phase methanol to water evolve in parallel during the irradiation, as for the richer mixture, showing that the sputtering reflects a direct fraction of the ice abundance evolution of these species. The ratio seems slightly higher than for the mixture with equal proportions of methanol and water in the ice. This measurement suggests that with a lower abundance of methanol with respect to water ice, more intact molecules are sputtered. 
However, based on a single mass m=31 and given the low signal-to-noise ratio of the QMS measurement, the QMS uncertainty on this methanol ice mixture is high and needs to be better constrained by additional measurements in the future.
%
%
%____________________________________________
\begin{figure}[tbhp]
\includegraphics[width=\linewidth]{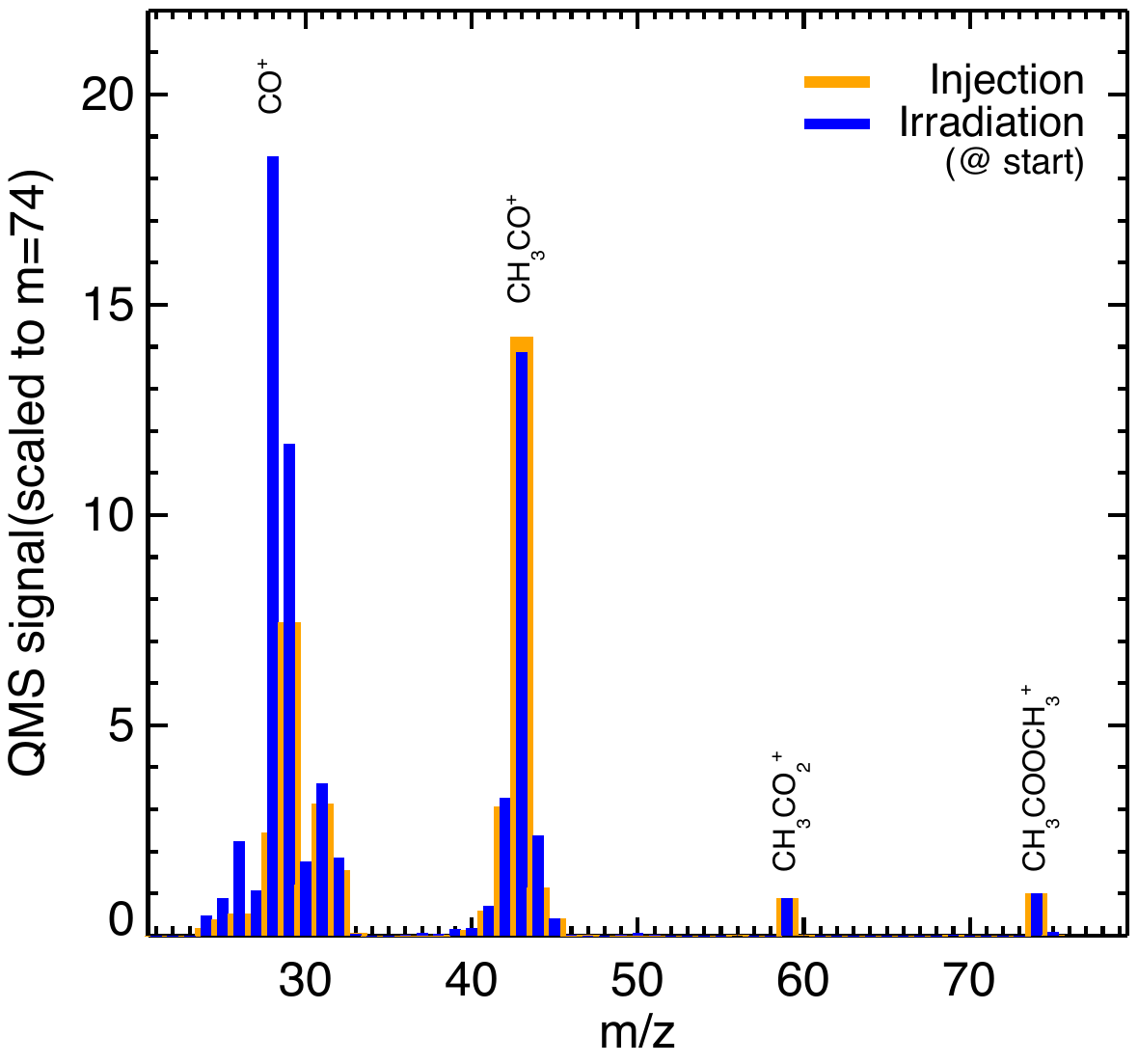}
\caption{Average of the mass spectra that were recorded during the methyl acetate injection to form the ice film (orange histogram), which trace the QMS mass fragmentation pattern of pure methyl acetate. We also show the average of a few spectra that were recorded at the beginning of the ice-film irradiation (blue histogram). Both mass spectra are scaled to mass 74, which corresponds to ionised methyl acetate for comparison. This shows that masses 74, 59, and 43 can be used to monitor methyl acetate because the radiolysis with the ion produces most contributions to peaks at lower masses. See text for details.} \label{fig:fragmentation_patterns_29W2_BIS}
\end{figure}
%____________________________________________
%
%____________________________________________
\begin{figure*}%[tbhp]
\begin{minipage}[c]{1.1\columnwidth}
\includegraphics[width=\linewidth]{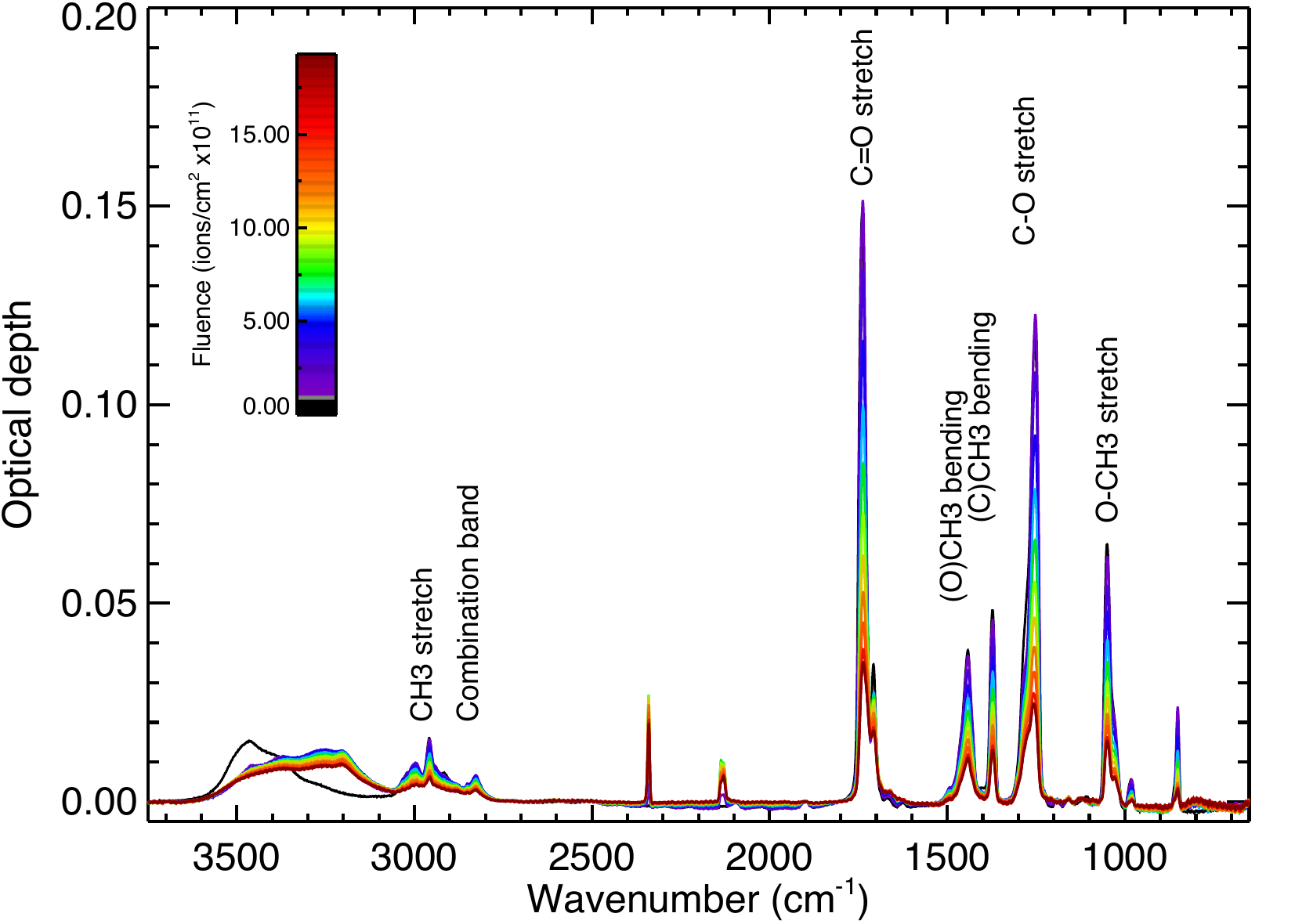}
\caption{Left: CH$_3$COOCH$_3$ ice-film experiment. Top panel: Ice column density measurements from IR spectra.
Right: Top panel: Ice column density measurements from IR spectra. 
Middle panel: QMS-scaled intensities at selected masses we used to follow carbon dioxide (m=44), methyl acetate (m=74,59,43,29,15), and CO (m=28) during irradiation. 
Lower panel: QMS signal ratios of the m=43 (CH$_3$CO) and m=59 (CH$_3$COO) biggest fragments to the m=74 (ionised CH3COOCH3) are shown as a function of irradiation (full lines). The ratios stay reasonably constant, with a value close to the pre-irradiation value (dotted lines). See text for details. \label{fig:mosaic_29W2_BIS}}
\end{minipage}
\begin{minipage}[c]{0.05\columnwidth}
\end{minipage}
\begin{minipage}[c]{0.85\columnwidth}
\includegraphics[width=\linewidth]{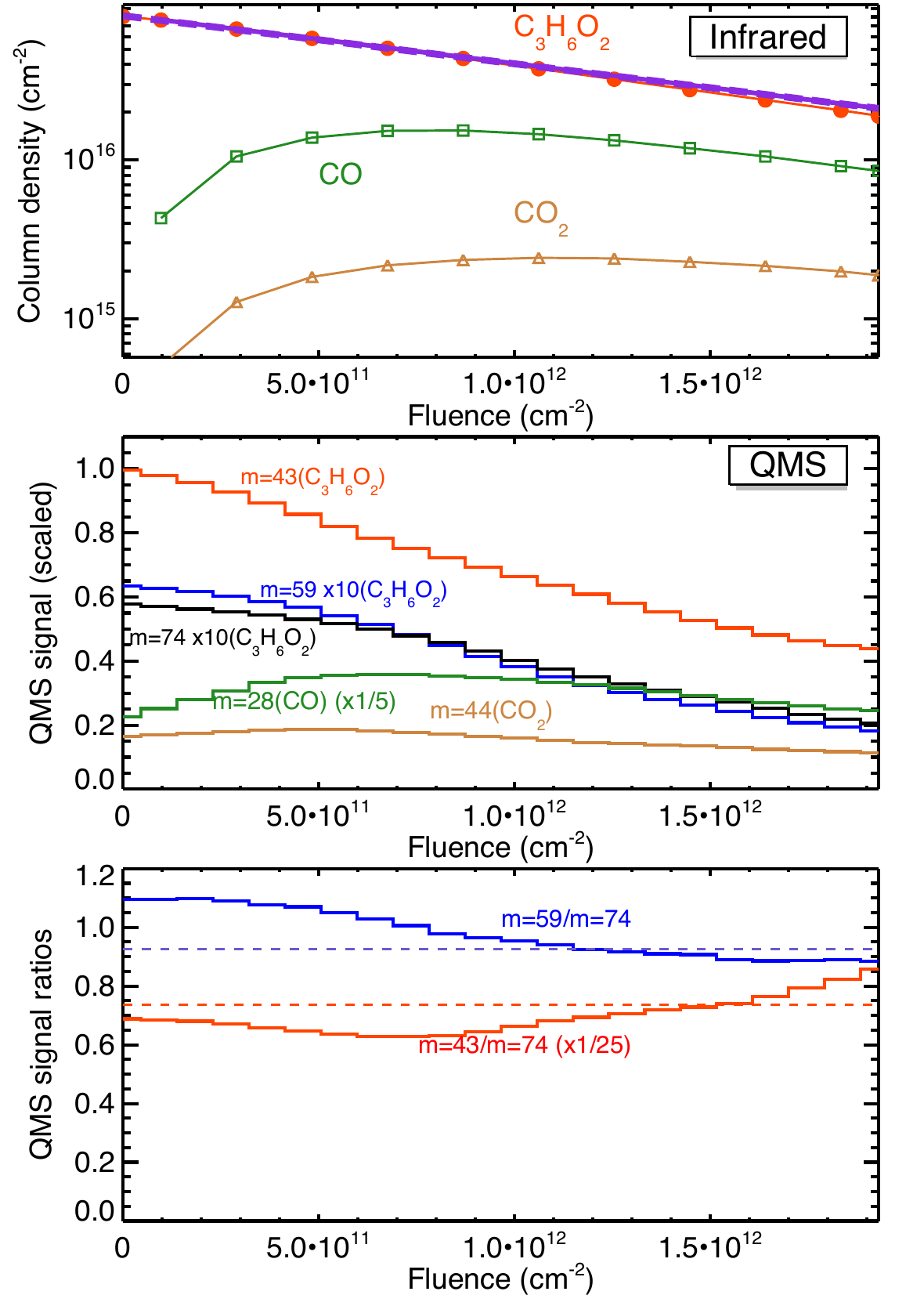}
\end{minipage}
\end{figure*}
%____________________________________________
%
%____________________________________________
\begin{figure*}%[tbhp]
\begin{minipage}[c]{1.1\columnwidth}
\includegraphics[width=\linewidth]{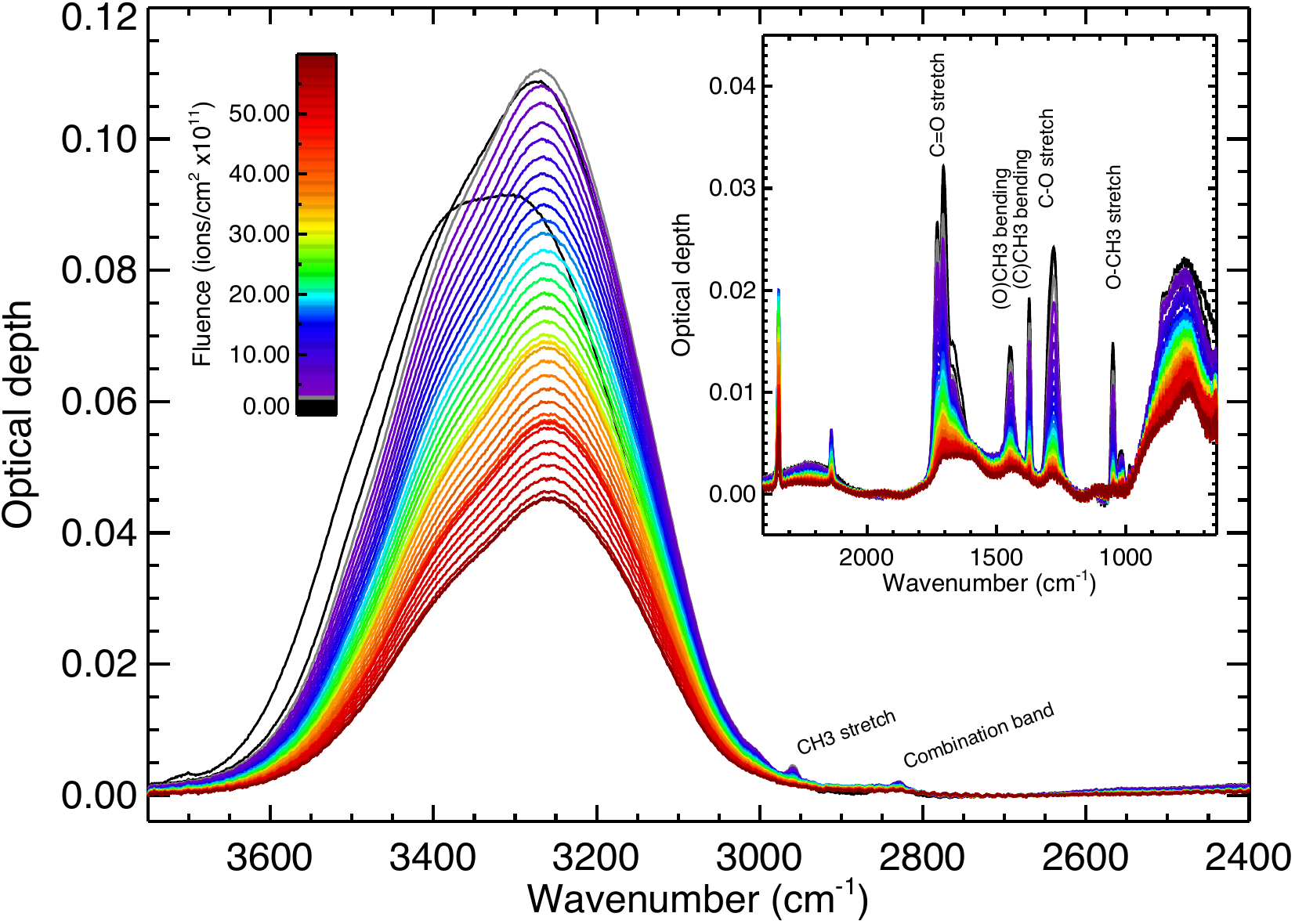}
\caption{Left: H$_2$O:CH$_3$COOCH$_3$ 100:13 ice-film experiment. Right: Top panel: Ice column density measurements from IR spectra. The black dashed lines represent the fits to the destruction cross section of water and CH$_3$COOCH$_3$.
Middle panel: QMS-normalised masses we used to follow water, methanol, and m=28 (CO, H$_2$CO, and CO$_2$) upon irradiation. 
m=28 increases at the beginning because it is mainly composed of fragments from the radiolytic products of the initial ice mixture.
Lower panel: CH$_3$COOCH$_3$/H$_2$O abundance ratio deduced from the IR spectra of the ice film as a function of fluence (red dots). Comparison with the QMS-determined abundance ratio of the same desorbed molecules (dark blue line). The blue and cyan star indicates the gas ratio measurement with the QMS during the ice-film deposition. \label{fig:mosaic_29W1_BIS}}
\end{minipage}
\begin{minipage}[c]{0.05\columnwidth}
\end{minipage}
\begin{minipage}[c]{0.85\columnwidth}
\includegraphics[width=\linewidth]{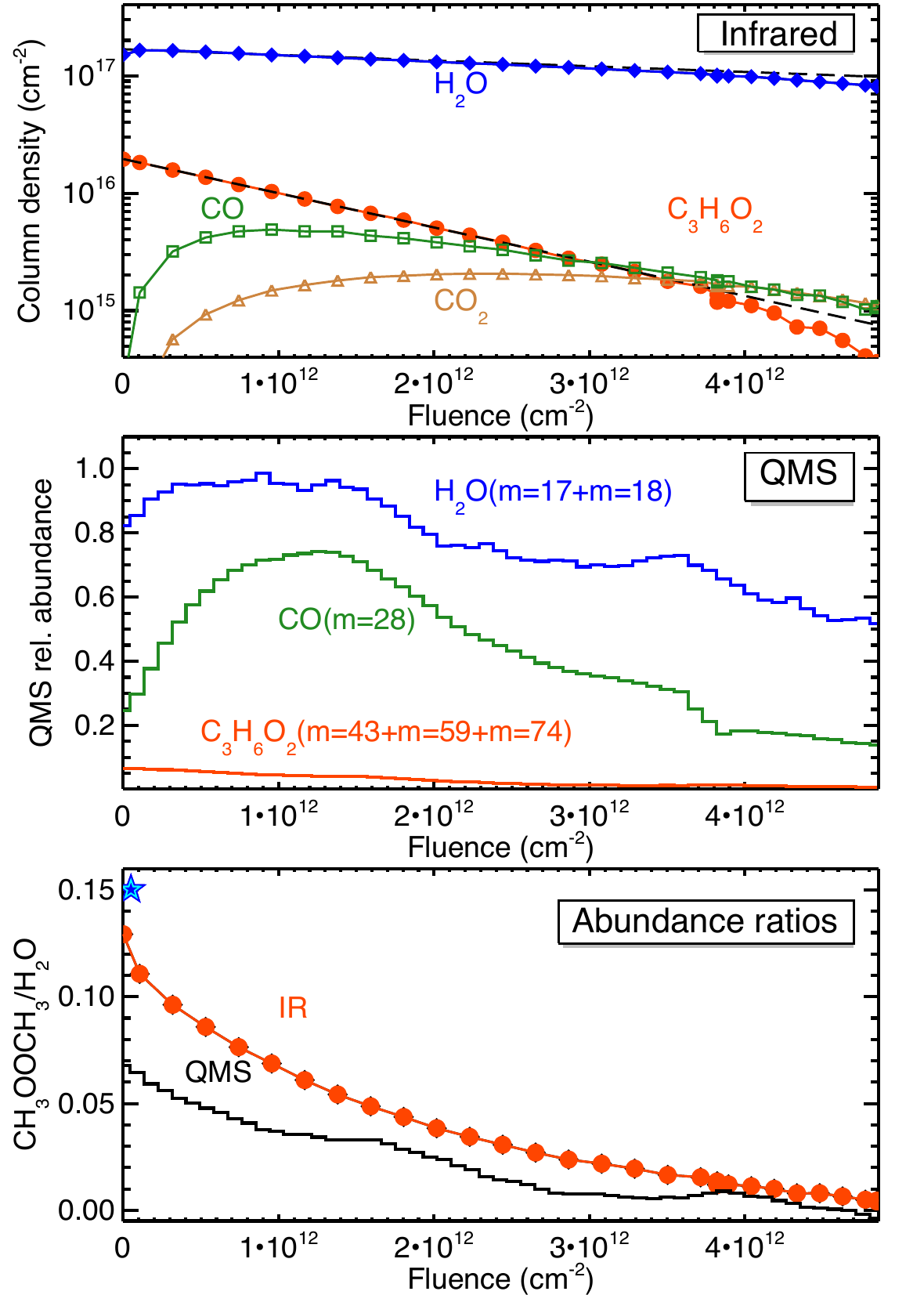}
\end{minipage}
\end{figure*}
%____________________________________________
%

\subsection{Radiolysis versus sputtering \label{section_radiolyse_versus_sputtering}}
For semi-infinite thick ice films, the evolution with the fluence (F) of the species column densities (N) in the ice film (monitored with the IR measurements) satisfies the following differential equation:
\begin{equation}
\rm \frac{dN_i}{dF}= -\sigma_{i}^{destruction}~N_i -Y_{effective}^{sputtering}~f_i + \sum_{j} \sigma_{j\rightarrow i}^{formation}~N_j
\label{equation_generale}
\end{equation}
for species i, where $\rm \sigma_{i}^{destruction}$ is the radiolytic destruction cross section. $\rm Y_{effective}^{sputtering}$ is the semi-infinite thickness effective-sputtering yield for the ice mixture we considered. It corresponds to the total number of sputtered molecules per ion, including the intact and the radiolysed ions. $\rm f_i$ is the fraction of species i in the ice surface layers that are involved in the sputtering, and $\rm \sigma_{j\rightarrow i}^{formation}$ is the induced radiolytic production of species i from another species j in the ice film.

For $\rm i=CH_3OH$, the induced radiolytic production from the daughter products is of lesser importance, and Eq.~\ref{equation_generale} can be approximated by
\begin{equation}
\rm \frac{dN_{CH_3OH}}{dF} \approx -\sigma_{CH_3OH}^{destruction}~N_{CH_3OH} -Y_{effective}^{sputtering}~f_{CH_3OH} 
\label{equation_reduite}
,\end{equation}
where $\rm f_{CH_3OH}$ is the fraction of methanol molecules in the sputtered ice layers (i.e. the fraction in the ice composition).\\
Adopting a cylindrical geometry for the sputtered volume \citep[e.g.][Fig.1]{Dartois2018}, we define $\rm N^d$ as the column density of molecules in the ice film that corresponds to the depth of desorption (i.e. the height of the cylinder). For methanol molecules mixed in a water matrix, $\rm N^d_{CH_3OH}\approx N^d_{H_2O} \times f_{CH_3OH}$.
The sputtering efficiency, that is, the observed yield in the electronic regime of stopping power, therefore depends on the fraction of molecules radiolytically processed
%\LEt{I cannot make out what you wish to say here or which word belongs to which other word. Please rephrase for clarity. Do you mean "the fraction of molecules that are radiolytically processed"?} 
during the ion interaction within this volume of ice with respect to the net amount of molecules contained in this ("ejected") volume.
We consider the product $\rm R^d_{CH_3OH}=\sigma_{CH_3OH}^{destruction}~N^d_{CH_3OH}$.\\
It is obvious that if $\rm R^d_{CH_3OH}$ approaches the value of $\rm Y_{effective}^{sputtering} \times f_{CH_3OH}$, most sputtered species should come out as fragments from the radiolysis.\\
The QMS abundance ratio measurements, which allow us to access the fraction of intact sputtered molecules $\rm \chi$ in the gas phase, should reflect this, and to first order,
\begin{align}
\rm \chi = \left(\frac{CH_3OH}{H_2O}\right)^{QMS}       & \rm \approx \frac{(Y_{effective}^{sputtering}-\sigma_{CH_3OH}^{destruction}~N^d) {\scriptstyle\times} f_{CH_3OH}}{(Y_{effective}^{sputtering}-\sigma_{H_2O}^{destruction}~N^d) {\scriptstyle\times} f_{H_2O.}} \label{equation_QMS}
\end{align}
The radiolytic destruction cross sections $\rm \sigma^{destruction}_{H_2O}$ of water and $\rm \sigma^{destruction}_{CH_3OH}$ of methanol were obtained to first order by fitting the IR measurement when the film is thick enough for the sputtering to not contribute significantly to the IR evolution.
Equation \ref{equation_QMS} may overestimate the radiolysis because the processes at the surface are not fully symmetrical with respect to the surface interface (e.g. secondary electrons produced close to the surface may escape before to act radiolytically on neihbouring species).

For the ice-mixture experiment shown in Fig.\ref{fig:mosaic_29W3_BIS}, the fitted CH$_3$OH cross section is $\rm \sigma^{destruction}_{CH_3OH} \approx  4.41\pm0.06\times10^{-13}$~cm$\rm^2/$ion (i.e. close to the pure CH$_3$OH cross section of $\rm \sigma^{destruction}_{CH_3OH} \approx  5.45\pm1.5\times10^{-13}$~cm$\rm^2/$ion fitted for the pure CH$_3$OH irradiations shown in Fig.~\ref{fig:mosaic_29W3_28_W2_BIS_28_W3_BIS}). The $\rm \sigma^{destruction}_{H_2O} \approx  9.8\pm0.3\times10^{-14}$~cm$\rm^2/$ion. The two fitted cross-section curves are shown overlaid as dashed lines in the right upper panel of Fig.\ref{fig:mosaic_29W3_BIS}.

In the case of a water-rich ice mantle, the sputtering yield should be close to the yield of water ice. The water-ice sputtering yield is $\rm 2\pm0.2\times10^{4}$ for this ion. The number of sputtered layers at this energy is about 30 ice monolayers, which corresponds to a column density of $\rm N_d\approx3\pm0.6\times10^{16} cm^{-2}$ \citep[][]{Dartois2018}. 

For this experiment, Eq.~\ref{equation_QMS} reads
\begin{align}
\rm \chi & \rm \approx  \frac{2\pm0.2{\scriptstyle\times}10^{4} -4.41\pm0.11{\scriptstyle\times}10^{-13}\times 3\pm0.6{\scriptstyle\times}10^{16}} 
{ 2\pm0.2{\scriptstyle\times}10^{4} -9.80\pm0.53{\scriptstyle\times}10^{-14}\times 3\pm0.6{\scriptstyle\times}10^{16}} {\scriptstyle\times} \left(\frac{f_{CH_3OH}}{f_{H_2O}}\right)_{ice} \nonumber \\
& \rm \approx 0.40\pm0.23 \times \left(\frac{f_{CH_3OH}}{f_{H_2O}}\right)_{ice.} \label{equation_QMS_29W3_BIS}
\end{align}
The expected percentage of intact methanol molecules that are ejected for this experiment is about 40\%. The calculated evolution with time is represented by the blue dashed line in the right lower panel of Fig.\ref{fig:mosaic_29W3_BIS}, compared to the QMS measurement, which is shown as the solid blue line.\\

For the ice-mixture experiment shown in Fig.\ref{fig:mosaic_28W3}, the fitted CH$_3$OH cross section is $\rm \sigma^{destruction}_{CH_3OH} \approx  5.16\pm0.39\times10^{-13}$~cm$\rm^2/$ion (i.e. close to the pure CH$_3$OH cross section of $\rm \sigma^{destruction}_{CH_3OH} \approx  5.45\pm1.5\times10^{-13}$~cm$\rm^2/$ion fitted for the pure CH$_3$OH irradiations shown in Fig.~\ref{fig:mosaic_29W3_28_W2_BIS_28_W3_BIS}). The $\rm \sigma^{destruction}_{H_2O} \approx  12.1\pm0.4\times10^{-14}$~cm$\rm^2/$ion. The two fitted cross-section curves are shown overlaid as dashed lines in Fig.\ref{fig:mosaic_28W3}.

For this experiment, Eq.~\ref{equation_QMS} reads 
\begin{align}
\rm \chi & \rm \approx  \frac{ 2\pm0.2{\scriptstyle\times}10^{4} -5.16\pm0.39{\scriptstyle\times}10^{-13}\times 3\pm0.6{\scriptstyle\times}10^{16} } 
{ 2\pm0.2{\scriptstyle\times}10^{4} -12.1\pm0.4{\scriptstyle\times}10^{-14}\times 3\pm0.6{\scriptstyle\times}10^{16}} {\scriptstyle\times} \left(\frac{f_{CH_3OH}}{f_{H_2O}}\right)_{ice} \nonumber \\
& \rm \approx 0.28\pm0.18 \times \left(\frac{f_{CH_3OH}}{f_{H_2O}}\right)_{ice.} \label{equation_QMS_28W3}
\end{align}

The expected percentage of intact methanol molecules that are ejected for this experiment is about 30\% with respect to the bulk ice composition, with a higher relative uncertainty than for the previous experiment. The calculated evolution with irradiation time is represented by the blue dashed line in the right lower panel of Fig.\ref{fig:mosaic_28W3}. The QMS measurements are above this calculation, closer to 65\%, which would mean that more methanol is preserved when it is less abundant and more diluted in the ice matrix.
\subsection{CH$_3$COOCH$_3$}
In order to explore the sputtering for an even larger COM, we performed an irradiation experiment with methyl acetate.
The evolution of the C$_2$H$_6$O$_2$ IR spectra, ice column densities, and QMS intensities as a function of fluence are displayed in Fig.\ref{fig:mosaic_29W2_BIS}.
As for the methanol case, this pure measurement allows us to fit the bulk destruction cross section of methyl acetate from IR data for the considered ion, yielding $\rm \sigma^{destruction}_{C_2H_6O_2}=7\pm1.5\times10^{-13}~cm^2/ion$.
We show in Fig. 5 the mass spectra recorded for methyl acetate (during the injection to form the ice film), which traces the QMS mass fragmentation pattern of pure methyl acetate, and the comparison with the average of a few spectra that were recorded at the beginning of the ice-film irradiation. The two mass spectra are scaled to mass 74, which corresponds to ionised methyl acetate, for comparison. This shows that masses 74, 59, and 43 can be used to monitor methyl acetate because the radiolysis with the ion mostly contributes to peaks at lower masses.
The different QMS signals for the main mass pattern-fragments at masses 74, 59, and 43 for the experiment are displayed in the middle right panel of Fig.\ref{fig:mosaic_29W2_BIS}. The ratio between the largest fragment signals at masses 59 (CH$_3$COO) and 43 (CH$_3$CO) to the pure ionised species recorded at mass 74 are displayed as a function of irradiation (full lines) in the lower panel. The value measured at injection is added for comparison (dashed lines).
The mass fragmentation pattern during irradiation for these higher mass fragments evolves close to the pattern that is measured during the gas-phase injection (i.e. without irradiation), thus showing that mostly intact CH$_3$COOCH$_3$ molecules dominate for these signals.
\subsection{CH$_3$COOCH$_3$:H$_2$O}
The methyl acetate mixed experiment was performed on a water-rich mixture, an H$_2$O:CH$_3$COOCH$_3$~(100:13) ice film. This mixture is dominated by water ice and was chosen for the same reasons as explained previously for methanol, in particular, because we expect such large COMs to be minor components with respect to water ice (and this mixture is already quite rich for such a large COM in astrophysics).
The recorded IR spectra and their evolution as a function of the ion fluence are shown in Fig.~\ref{fig:mosaic_29W1_BIS}. 
We report the column density of these molecules in the ice using a C-O stretching mode of methyl acetate and the OH stretch of water ice.
 The selected QMS mass intensities are shown, which allows us to monitor the gas-phase evolution of water (m=17+m=18) and methyl acetate (m=74, 59, and 43 masses are identified as stable in the methyl acetate fragmentation pattern during deposition and at the beginning of irradiation).
Mass m=28 was also followed to monitor the contribution of CO (main contributing species that also contains less important contributions from the CO$_2$, H$_2$CO, CH$_3$OH fragmentation pattern).
The destruction cross section of CH$_3$COOCH$_3$ obtained for this mixture from the IR is evaluated to $\rm \sigma^{destruction}_{C_2H_6O_2}=6.7\pm1{\scriptstyle\times}10^{-13}~cm^2/ion$.
The observed QMS abundance ratio of CH$_3$COOCH$_3$/H$_2$O in experiment 29W1BIS (Fig.\ref{fig:mosaic_29W1_BIS}, panel with the abundance ratios) evolves in parallel with the species ratio of the ice film that is observed in the IR at any time of the irradiation. The bulk composition of the ice evolves with the fluence because of the radiolytic processes that occur inside, but the sputtering, which is a surface process, probes the ice composition.
%
%____________________________________________
\begin{figure*}%[tbhp]
\centering
\includegraphics[width=\linewidth]{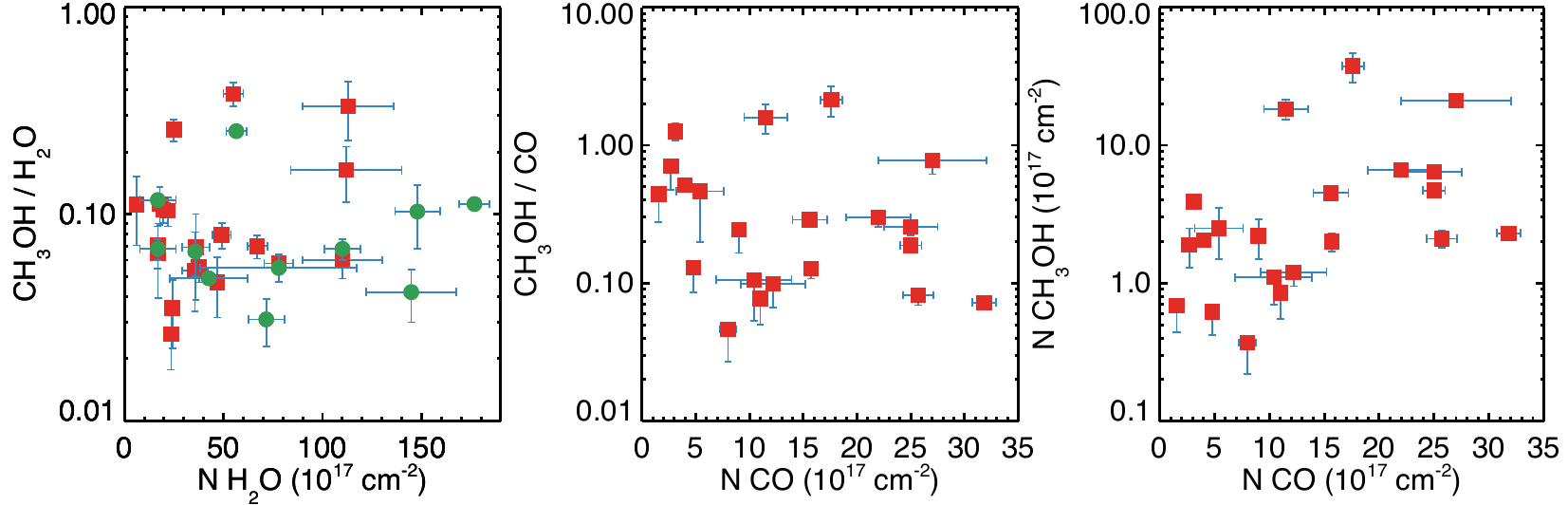}
\caption{Correlation plots for the ice column density for CH$_3$OH/H$_2$O vs. H$_2$O, CH$_3$OH/CO vs. CO, and the CH$_3$OH vs. CO for the sources presented in
%\LEt{please check your LaTeX commands for the two references } 
\citet[][red squares]{Whittet2011} and \citet[][green circles]{Bottinelli2010}.}
\label{fig:methanol_astro}
\end{figure*}
%____________________________________________
%
%
\section{Discussion}
\label{discussion}

\subsection{Methanol in interstellar ices}
Observationally, no strong generic correlation exists between the column densities of CO ice and CH$_3$OH \citep[e.g.][]{Whittet2011}.
Because of this lack of correlation, it is difficult if not problematic for an observer to predict a priori the observed methanol column density from the observed ice-mantle water, carbon dioxide, and carbon monoxide column densities for a given line of sight. 
This absence of a strong correlation is shown in Fig.\ref{fig:methanol_astro}, where we report the column densities of CH$_3$OH/H$_2$O versus H$_2$O, CH$_3$OH/CO versus CO, and the CH$_3$OH versus CO for sources investigated in \cite{Whittet2011}. After analysing this large dataset of CH$_3$OH observations, the authors conclude that its concentration in the ices towards embedded stars show order-of-magnitude object-to-object variations.
The observations seem to show that the occurrence of methanol ice requires high column densities of matter, which generally corresponds to a (possibly anterior) very cold or deeply protected environment phase \citep[with A$\rm_V>15$, see Fig.7 of][]{Boogert2015}. This is a necessary but maybe not sufficient condition because some lines of sight show stringent upper limits even at A$\rm_V>30$. The launch of the James Webb Space Telescope (JWST) is expected to help push these upper limit constraints towards detections, to properly quantify source-to-source variations.

It has long been hypothesised that methanol ice forms from the direct hydrogenation of CO ice. In which phase the methanol is observed in interstellar ice is debated through comparisons to laboratory data \citep[e.g.,][and references therein]{Penteado2015, Cuppen2011}. Definite observational proof of the statement that this is the unique (or at least the by far dominant) pathway to methanol formation is still lacking. 
Furthermore, the observed interstellar profiles of the carbon dioxide bending mode in some lines of sight can be reproduced in the laboratory and explained by the formation of
methanol-carbon dioxide intermolecular complexes, which would suggest that CO$_2$ is also part of the methanol environment. The formation of such complexes is affected by a dominant fraction of water ice, in which case the water molecules control the structure of the ice and inhibit the formation of such complexes \citep[][]{Dartois1999, Ehrenfreund1999, Klotz2004}.

\begin{figure}%[tbhp]
\centering
\includegraphics[width=\linewidth]{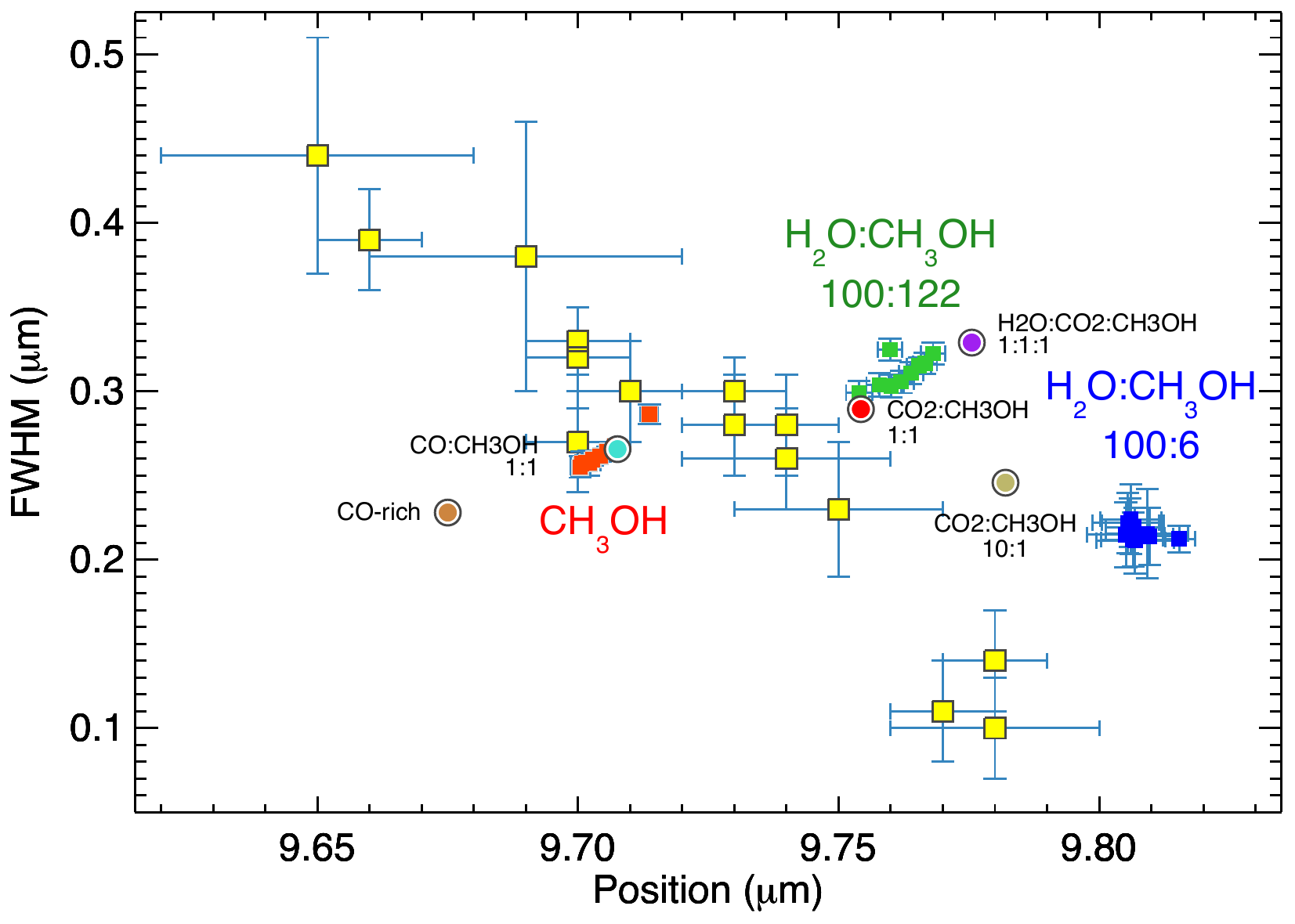}
\caption{Methanol C-O stretching-mode position-FWHM diagram of the experiments performed in this work, compared to ISM measurements presented in \cite{Bottinelli2010} (yellow squares), as well as the CO-rich point given in this article, and position FWHM from ice spectra of various compositions from the Leiden ice database, extracted in the same way as for our irradiation measurements.}
\label{fig:positions}
\end{figure}

CH$_3$OH has been investigated with Spitzer space telescope observations in dense regions \citep[][]{Bottinelli2010}.
Botinelli and coworkers show in their Fig.12 that most of the observed methanol C-O stretching-mode position versus full width at half-maximum (FWHM) is in a mantle whose composition is neither dominated by a H$_2$O-rich nor by a CO-rich matrix. The methanol C-O stretching-mode position versus FWHM diagram for our experiments is reported in such a diagram in Fig.\ref{fig:positions}, together with these previous measurements and other laboratory data. This diagram shows that different ice-mantle mixtures are still compatible with the positions we observed,  which encourages us to explore several alternatives to the ice-mantle composition in laboratory experiments.
%
%%%%%%%%%%%%%%%%%%%%%%%%%%%%%%%%%%%%%%%%%%%%%%%%%%%%%%%%%%%%%%%%%%%%%%%%%%%%%%%%
%____________________________________________
%____________________________________________
%

\subsection{Stopping-power destruction cross-section dependency}
\begin{figure}%[tbhp]
\centering
\includegraphics[width=\linewidth]{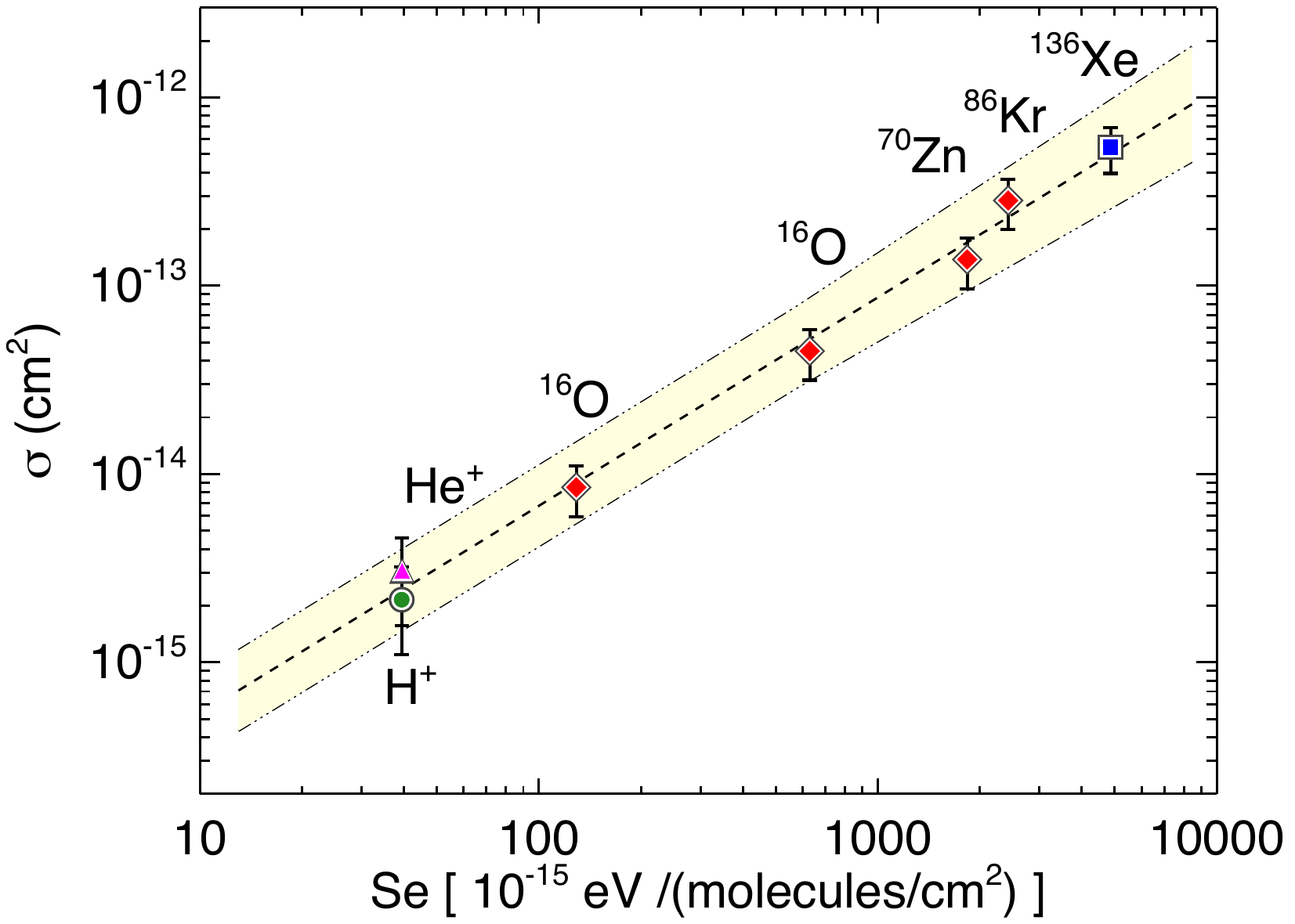}
\includegraphics[width=\linewidth]{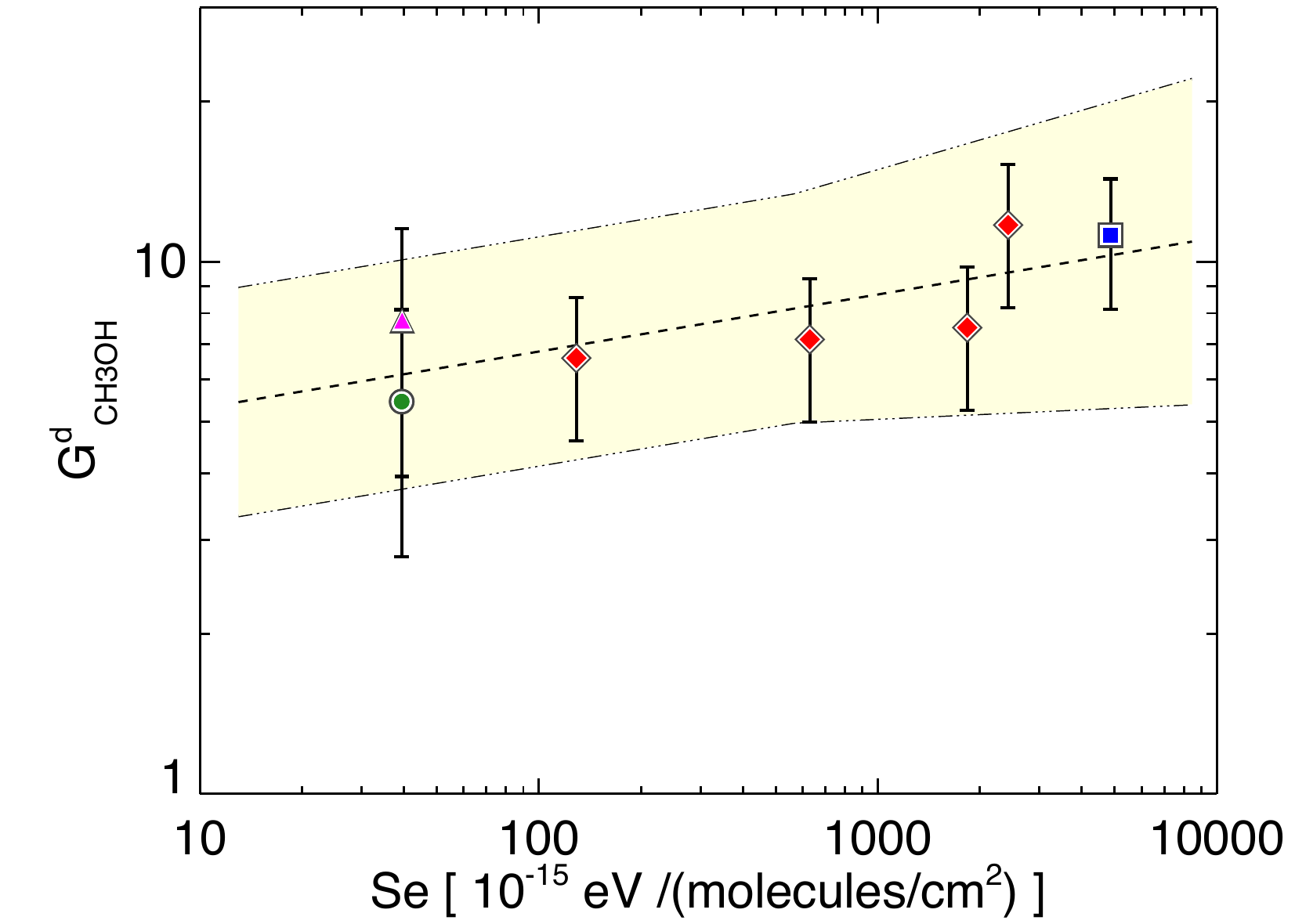}
\caption{Upper panel: Methanol destruction cross section as a function of stopping power in the electron regime. Red diamonds show data from \cite{Barros2011}, the magenta triangle represents 30~keV He$^+$ \citep{Baratta2002}, the green circle shows 200~keV H$^+$ \citep{Brunetto2005}, and the blue square shows data from this work. The dashed line represent the best least-squares fit to the cross section, and the filled region delineates the 95\% interval confidence. See text for details.
Lower panel: Corresponding methanol radiochemical destruction yield.}
\label{fig:ch3oh_destruction}
\end{figure}

%
%\LEt{please consider rephrasing so that you don't have four nouns that qualify the fifth noun, this is really awkward}
%Methanol-ice 
Destruction cross sections for methanol ice have been reported for different stopping powers \citep[Fig.8,][and references therein]{Barros2011}. The dependency of this ice on the electronic stopping power follows a power law, $\rm \sigma \propto Se^{n}$. We used the data of this article and reanalysed some of the data.
We excluded from the analysis the previously included 3~keV He$^+$ measurement for which the nuclear stopping power (Sn) is of the same order of magnitude as the electron stopping power (Se, this point therefore mixes two different regimes), as well as the proton irradiation at 0.8~MeV value \citep{Gerakines2001} because the cross section was not measured for a pure CH$_3$OH ice but in a mixture that also contained H$_2$O and CO$_2$.
We reanalysed the data reported in \cite{Baratta2002} for 30~keV He$^+$ (because the authors did not provide a destruction cross section) by fitting the data presented in Fig.7. This provided a cross section $\rm \sigma^{destruction}_{CH_3OH}=3.06\pm1.55\times10^{-15}~cm^2/ion$. We used the stopping power in the electron regime calculated for this ion using the SRIM software package 
%\LEt{please introduce}
 \citep[][The stopping and range of ions in matter]{Ziegler2010}, leading to $\rm 39.5\times10^{-15} eV molecules^{-1} cm^{2}$.
We added the mean destruction cross section for the 16K measurements of two methanol near-IR transitions with a 200~keV H$^+$ projectile by \cite{Brunetto2005}, which provided a mean cross section of $\rm \sigma^{destruction}_{CH_3OH}=2.15\pm1.05\times10^{-15}~cm^2/ion$. We used the electron stopping power calculated with the SRIM package, that is, $\rm 39.4\times10^{-15} eV molecules^{-1} cm^{2}$.
The resulting regression fit on the data gives a dependency of $\rm \sigma^{destruction}_{CH_3OH}=4.14\pm2.0\times10^{-17} Se^{1.11\pm0.11}$, which we show in the upper panel of Fig.\ref{fig:ch3oh_destruction}.
The methanol destruction radiochemical yield $\rm G^d_{CH_3OH}=100{\scriptstyle\times}\sigma^{destruction}_{CH_3OH}/Se$, that is, the number of methanol molecules that are destroyed by radiation per 100 eV of absorbed energy, and its evolution, can be derived from the previous fit. We show this in the lower panel of Fig.\ref{fig:ch3oh_destruction}. Its evolution shows that at lower stopping power in the electron regime, a higher fraction of methanol is preserved, and the fraction of mean sputtered intact molecules integrated over the cosmic-ray distribution should therefore be higher.
%
%%%%%%%%%%%%%%%%%%%%%%%%%%%%%%%%%%%%%%%%%%%%%%%%%%%%%%%%%
%
\subsection{Radiolytic destruction versus sputtering dependency as a function of energy and stopping power}

The dependency of our  results on the stopping power is an important question for a large distribution of energies in cosmic-ray particles.
The relevant astrophysical criterion is the fraction of intact molecules that are sputtered from an interstellar ice mantle, or in other words, the radiolytic to sputtering efficiency as a function of energy.
As discussed above, to first order, the number of molecules that are dissociated in the sputtered volume for a species X is
\begin{align}
\rm R^d_{X}              \approx \sigma^{destruction}_{X}\times N^d \times f_{X} = \sigma_{0} S_e^n \times N^d_0 Se^m \times f_{X}, \nonumber
\end{align}
where $\rm N^d$ is the column density of molecules in the ice film, which corresponds to the depth of the desorption when we assume that all the molecules come from the same sputtered volume. $\rm \sigma_{0} Se^n$ is the dependency of the destruction cross section with the stopping power in the electron regime, as shown and discussed in the previous section. We also expect the depth dependency to follow a power law, with a power-law index m.
The sputtered molecule yield is
\begin{align}
\rm Y_{eff} = S\times N^d  = S_0 S_e^p \times N^d_0 Se^m = Y_0 Se^{p+m},\nonumber% molecules = cm2 * molecules/cm^2 
\end{align}
where $\rm S$ is the surface (cm$^2$) of a sputtering "cylinder".

The fraction of dissociated to total sputtered molecules in the sputtered volume is then
\begin{align}
\rm \eta = \frac{R^d_{X}}{Y_{eff}} \approx \frac{\sigma^{destruction}_{X}\times N^d \times f_{X}}{S\times N^d} =  \frac{\sigma_{0} Se^n \times N^d_0 Se^m \times f_{X}}{S_0 Se^p \times N^d_0 Se^m,}
\end{align}
where the dependency of the different parameters with the stopping power in the electronic regime is specified. Thus,
\begin{align}
\rm \eta = \frac{R^d_{X}}{Y_{eff}} = C_{0} Se^{n-p} \times f_{X},
\end{align}
where $\rm C_{0}$ is a constant.
From previous measurements, we know that the sputtering yield evolves quadratically with respect to the stopping power in the electron regime for a large majority of molecular cryogenic solids \citep[e.g. H$_2$O, CO$_2$, and CO,][]{Rothard2017}, thus $\rm p+m\approx2$, then $\rm p\approx2-m$.

In general, the measured and adjusted destruction cross sections for ices at low temperature show a power-law dependency with $\rm 1.<n<1.5$, for example, CH$_3$OH, n=1.1, this work; CO$_2$ \citep[n=1.1, Fig.5,][]{Mejia2015} as a function of the electronic stopping power. 
As discussed in \cite{Dartois2018}, experiments and thermal spike models of the phase transformation induced by the ion track in insulators predict a dependency of the radius r of the cross section that evolves as $\rm r \sim \sqrt{S_e}$, where $\rm S_e=dE/dx$ is the deposited energy per unit path length \citep[e.g.][]{Lang2015,Toulemonde2000,Szenes1997}, and with a threshold in $\rm S_e^{th}$ to be determined. In this case, this implies that $\rm m\approx1$. 
When we combine these arguments, we find
\begin{align}
\rm \eta = \frac{R^d_{X}}{Y_{eff}} = C_{0} Se^{[0-0.5]} \times f_{X.}
\end{align}
Within this regime, the fraction of dissociated molecules in the sputtered volume will therefore not strongly depend on the stopping power and will decrease at lower stopping power in the electron regime, as shown in the lower panel of Fig.\ref{fig:ch3oh_destruction}. The majority of the molecules, even for COMs, should be sputtered intact, in contrast to VUV photodesorption. 
\subsection{Sputtering rates}
\subsection*{Methanol in water ice}
\begin{figure}%[tbhp]
\centering
\includegraphics[width=\linewidth]{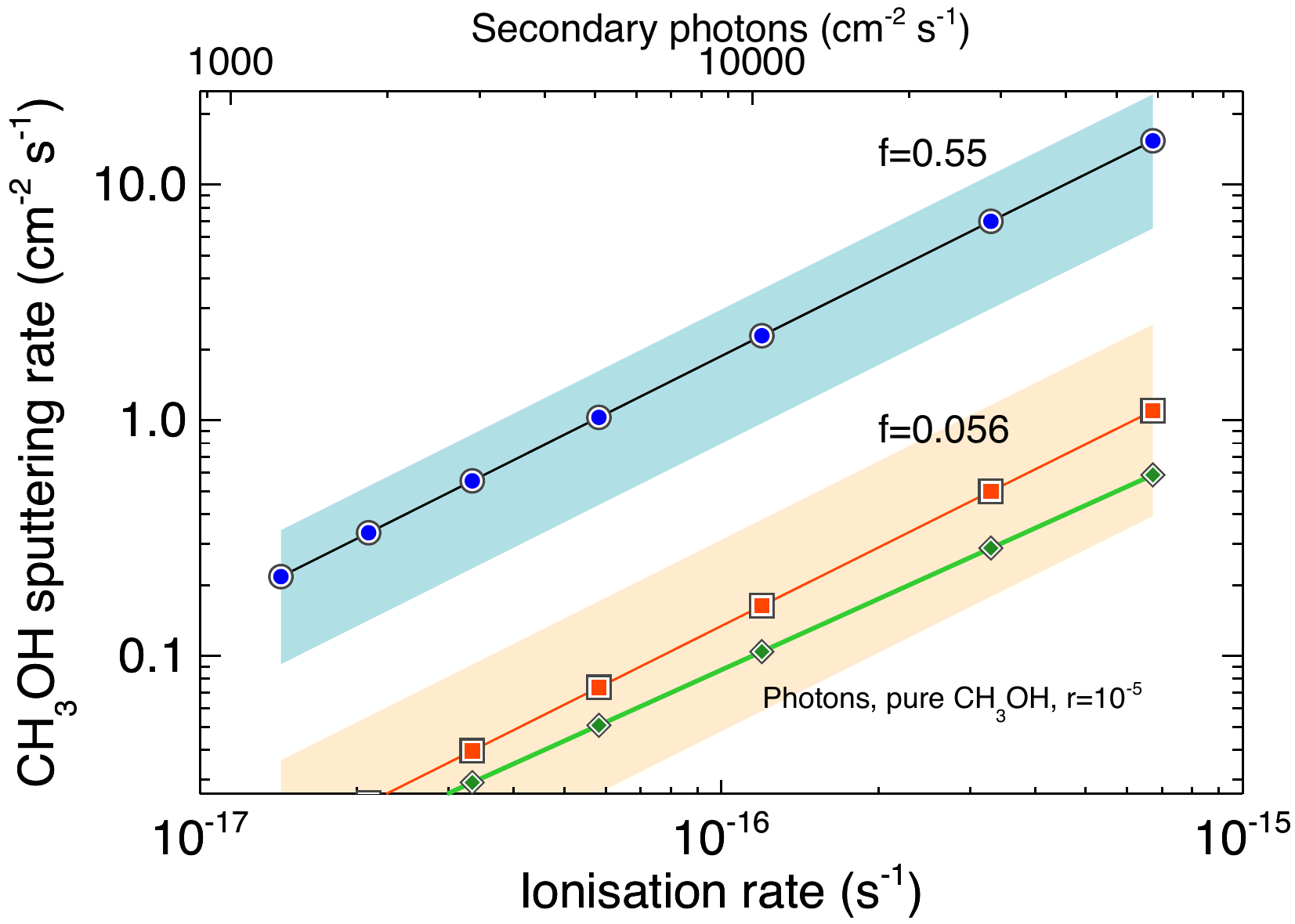}
\caption{Methanol in a water-ice matrix sputtering rate as a function of the cosmic-ray ionisation rate. The different curves correspond to the methanol fraction with respect to the water ice that we measured here. This range covers most of the observational dispersion. This sputtering rate must formally be multiplied by two because the cosmic-ray projectiles move through the interstellar grains.
The upper axis gives an estimate of the evolution of the corresponding secondary VUV photon flux \citep[e.g.][]{Shen2004, Prasad1983}, given for comparison with photodesortion rates. The rate corresponding to $\rm \sim~10^{-5} CH_3OH/VUV~photon$ for a pure methanol-ice film is shown in green. This is the order of magnitude measured by \cite{Bertin2016}, whereas \cite{Cruz-Diaz2016} obtained an upper limit that is three times higher. Cosmic-ray sputtering rates for methanol that is embedded in a CO matrix would be orders of magnitude higher because CO desorbs more efficiently.}
\label{fig:ch3oh_rate}
\end{figure}

Assuming that the fraction of dissociated molecules does not depend on the stopping power, we calculated the astrophysical sputtering rate in the same way as discussed in \cite{Dartois2015}, using Eq.~\ref{equation_QMS}. The result is shown in Fig.\ref{fig:ch3oh_rate} for the measured fractional methanol abundances in a water-dominated ice mantle as a function of different ionisation rates. The filled regions give the extent of the rate, including the measured uncertainties.
We added to this figure an estimate of the relation between the expected secondary VUV photon flux and the ionisation rate, as discussed in \cite{Shen2004}, under dense-cloud conditions when the external UV field is fully attenuated. We show a photodesorption rate of $\rm \sim10^{-5}$/VUV photon, which approximately corresponds to the estimated pure methanol-ice desorption in \cite{Bertin2016}, whereas \cite{Cruz-Diaz2016} obtained an upper limit of about $\rm <3\times10^{-5}$/VUV photon. 
\subsection*{Methanol in carbon monoxide ice}
When methanol is mixed with CO ice, the photodesorption rate is considerably lower than for pure methanol, with fewer than $\rm \sim10^{-6}$/VUV photon \citep{Bertin2016}. In thermal desorption studies of mixtures with a temperature-programmed desorption (TPD) technique, with heating rates of about 10K/min \citep[e.g.][]{Ligterink2018}, methanol does not desorb efficiently, while CO sublimates easily.
Unlike TPD experiments, the rate of the heating spike that is induced by cosmic rays may reach $\rm 10^{12}-10^{14}~K/s$, and most of the embedded methanol molecules may be driven by the sudden sputtering of the main ice constituents.
To estimate the expected sputtering magnitude if methanol were embedded in a CO matrix and driven by the thermal spike, we therefore replaced H$_2$O by CO in Eq.~\ref{equation_QMS}.
The effective yield for a pure CO ice can be estimated from the equation in Fig.6 of \cite{Seperuelo2010}, $\rm Y=0.033\times Se^2$, with an Se of about $\rm2920~10^{-15} eV/(CO/cm^2)$ which was estimated with the SRIM package for a density of CO ice of 0.8$\rm g/cm^3$ \citep{Roux1980, Bouilloud2015}. This led to about $2.8\times10^5$ CO/ion. In the absence of constraints and when we assume an aspect ratio close to 1 for the sputtered volume, this corresponds to about 66 CO layers, or about $\rm4.4\times10^{16}cm^{-2}$ in equivalent column density for the sputtered depth. The CO destruction cross section can be evaluated to lie between $\rm1.7~to~2.3\times10^{-13}cm^{2}$ from previous measurements when we assume a dependency between $\rm Se^1$ and $\rm Se^{3/2}$ based on the previous 50 MeV Ni experiment reported by \cite{Seperuelo2010}. As reported in Eq.~\ref{equation_QMS}, we expect 95\% of the bulk ice CH$_3$OH/CO to be sputtered intact. Even with an aspect ratio of 5, which favours methanol radiolysis more, the factor decreases only to 70\%. The global rates are therefore about a factor 40 higher than for a water-ice matrix.
Following the arguments we discussed above, the sputtering rates by cosmic rays for methanol that is embedded in a CO or a CO$_2$ matrix will be order(s) of magnitude higher than our measurements in water ice.\\
%
%____________________________________________
%____________________________________________
%
%
%____________________________________________
%____________________________________________
%
%
\subsection*{Larger species}
As discussed in section \ref{section_radiolyse_versus_sputtering}, if $\rm \sigma_{X}^{destruction}~N^d =Y_{effective}^{sputtering} $, most sputtered species are expected to come out as fragments from the radiolysis. The limiting destruction cross section from this equality and the ion we used in our experiments for a water rich mixture should be $\rm \sigma_{X}^{destruction} \sim 2\pm0.2{\scriptstyle\times}10^{4} / 3\pm0.6{\scriptstyle\times}10^{16} = 6.67\pm{\scriptstyle\times}10^{-13}~cm^2/ion$, which is close to the cross section that was evaluated for methyl acetate.
However, the measurements displayed in Fig.~\ref{fig:mosaic_29W1_BIS}  show that a large portion of CH$_3$COOCH$_3$ molecules are sputtered intact, that is, the radiolysis of molecules that are ejected from the surface seems less efficient. The sputtering process of intact molecules when the COM is diluted in the ice also seem higher than expected from Eq.~\ref{equation_QMS}.\\
Simple physical assumptions were made in the calculations we presented in section~\ref{section_radiolyse_versus_sputtering} for instance for the adopted constant effective sputtering yield, the constant depth of desorption for the mixtures, the cylindrical shape for the sputtered volume, or that the cylindric radius assumed for the sputtered volume is the same for the radiolysis radius. For the latter, it has all reasons to be different. The ejected fragments that are observed can occur in the direct excitation process in the core of the track (infratrack), or they may be partly the result of some of the secondary (delta-)electron processing, but the sputtering of intact molecules likely occurs from a much larger radius.
These assumptions are of first order and too simple to represent the complexity of these mixtures. % after the irradiation has begun.
The calculations presented in section~\ref{section_radiolyse_versus_sputtering} explain the order of magnitude of intact sputtered molecules. They also follow the decrease in measured ratio from QMS data of the gas phase for intact species that are sputtered upon irradiation, in agreement with the ice composition in the bulk, as monitored by the IR measurements. However, they probably underestimate the sputtering of intact species because even larger COMs molecules can be sputtered intact by swift heavy ions \citep[e.g. leucine, M=131~u, ergosterol, M=393~u, and even larger,][]{Hedin1987, Hakansson1982, Hakansson1989}, and calls for the development of a more advanced model. %, for methanol rich mixtures.}
\section{Conclusions}

We have measured the electron sputtering yield of ice mixtures including COMs that are observed in the solid phase or in the gas phase in the ISM by simulating methanol and methyl acetate embedded in water-rich ice mantles at 10K exposed to interstellar cosmic rays. We conclude that for COMs, whose abundance is generally a small fraction of the main ice-matrix component, a large portion of intact molecules are desorbed by cosmic rays and with a sputtering efficiency that is close to that of the ice matrix.\\
Our measurements were performed at a single electron stopping power, but because COMs are in most cases a minor component with respect to the water-ice dominating mantle, the measured quadratic water-ice sputtering-yield dependency on the stopping power can be used to extrapolate the effective yield in a distribution of cosmic rays in an astrophysical medium, as was previously done for pure water ice.
The investigation made in this article is a step in the development of a more systematic study of the sputtering efficiency of COMs (methanol and larger molecules) in various ice matrices that are relevant for interstellar contexts. The next experiments will probe mixtures of methanol-CO$_2$ and methanol-CO. Based on the experiments performed in this work, we expect the same qualitative behaviour, that is, that the sputtering yield is dominated by the effective yield of the matrix, which is higher (more efficient by up to orders of magnitude for CO) than for H$_2$O, and that a large portion will be ejected intact.
Measurement campaigns are also scheduled to explore the composition and depth dependency parameters in order to refine the energy dependency
%\LEt{you use both "dependence" and "dependency". Please decide for one and use it throughout to avoid confusion}
 of the parameters on this process.\\
The calculations indicate that cosmic-ray sputtering dominates secondary photons for CH$_3$OH desorption. This is most probably a general case for larger COMs as well.
\begin{acknowledgements}
This work was supported by the Programme National "Physique et Chimie du Milieu Interstellaire" (PCMI) of CNRS/INSU with INC/INP co-funded by CEA and CNES, by the P2IO LabEx program: "Evolution de la mati\`ere du milieu interstellaire aux exoplan\`etes avec le JWST" 
and the ANR IGLIAS, grant ANR-13-BS05-0004 of the French Agence Nationale de la Recherche.
Experiments performed at GANIL. We thank T. Madi, T. Been, J.-M. Ramillon, F. Ropars and P. Voivenel for their invaluable technical assistance. 
A.N. Agnihotri acknowledges funding from INSERM-INCA (Grant BIORAD) and R\'egion Normandie fonds Europ\'een de d\'eveloppement r\'egional-FEDER Programmation 2014-2020.
We would like to acknowledge the anonymous referee for very constructive comments that significantly improved the content of our article, as well as the editor M. Tafalla and language editor A. Peters. 
\end{acknowledgements}

%\nocite{*}
%-------------------------------------------------------------------


\begin{thebibliography}{}

%IGLIAS: A new experimental set-up for low temperature irradiation studies at large irradiation facilities
\bibitem[Aug{\'e} et al.(2018)]{Auge2018} Aug{\'e}, B., Been, T., Boduch, P., et al.\ 2018, Review of Scientific Instruments, 89, 075105 

%Detection of complex organic molecules in a prestellar core: a new challenge for astrochemical models
\bibitem[Bacmann et al.(2012)]{Bacmann2012} Bacmann, A., Taquet, V., Faure, A., Kahane, C., \& Ceccarelli, C.\ 2012, \aap, 541, L12 

%A comparison of ion irradiation and UV photolysis of CH4 and CH3OH
\bibitem[Baratta et al.(2002)]{Baratta2002} Baratta, G.~A., Leto, G., \& Palumbo, M.~E.\ 2002, \aap, 384, 343 

%Chemical and physical effects induced by heavy cosmic ray analogues on frozen methanol and water ice mixtures
%\bibitem[de Barros et al.(2014)]{Barros2014} de Barros, A.~L.~F., da Silveira, E.~F., Rothard, H., Langlinay, T., \& Boduch, P.\ 2014, \mnras, 443, 2733 

%Radiolysis of frozen methanol by heavy cosmic ray and energetic solar particle analogues
\bibitem[de Barros et al.(2011)]{Barros2011} de Barros, A.~L.~F., Domaracka, A., Andrade, D.~P.~P., et al.\ 2011, \mnras, 418, 1363.

%UV Photodesorption of Methanol in Pure and CO-rich Ices: Desorption Rates of the Intact Molecule and of the Photofragments
\bibitem[Bertin et al.(2016)]{Bertin2016} Bertin, M., Romanzin, C., Doronin, M., et al.\ 2016, \apjl, 817, L12 

%Observations of the icy universe.
\bibitem[Boogert et al.(2015)]{Boogert2015} Boogert, A.~C.~A., Gerakines, P.~A., \& Whittet, D.~C.~B.\ 2015, \araa, 53, 541 

%The c2d Spitzer Spectroscopic Survey of Ices Around Low-mass Young Stellar Objects. IV. NH3 and CH3OH
\bibitem[Bottinelli et al.(2010)]{Bottinelli2010} Bottinelli, S., Boogert, A.~C.~A., Bouwman, J., et al.\ 2010, \apj, 718, 1100 

%Bibliographic review and new measurements of the infrared band strengths of pure molecules at 25 K: H2O, CO$_2$, CO, CH4, NH3, CH3OH, HCOOH and H$_2$CO
\bibitem[Bouilloud et al.(2015)]{Bouilloud2015} Bouilloud, M., Fray, N., B{\'e}nilan, Y., et al.\ 2015, \mnras, 451, 2145 

%Reflectance and transmittance spectra (2.2 2.4 ?m) of ion irradiated frozen methanol
\bibitem[Brunetto et al.(2005)]{Brunetto2005} Brunetto, R., Baratta, G.~A., Domingo, M., \& Strazzulla, G.\ 2005, \icarus, 175, 226 

%Seeds Of Life In Space (SOLIS): The Organic Composition Diversity at 300-1000 au Scale in Solar-type Star-forming Regions
\bibitem[Ceccarelli et al.(2017)]{Ceccarelli2017} Ceccarelli, C., Caselli, P., Fontani, F., et al.\ 2017, \apj, 850, 176 

%Negligible photodesorption of methanol ice and active photon-induced desorption of its irradiation products
\bibitem[Cruz-Diaz et al.(2016)]{Cruz-Diaz2016} Cruz-Diaz, G.~A., Mart{\'{\i}}n-Dom{\'e}nech, R., Mu{\~n}oz Caro, G.~M., \& Chen, Y.-J.\ 2016, \aap, 592, A68 

%CO ice mixed with CH3OH: the answer to the non-detection of the 2152 cm-1 band?
\bibitem[Cuppen et al.(2011)]{Cuppen2011} Cuppen, H.~M., Penteado, E.~M., Isokoski, K., van der Marel, N., \& Linnartz, H.\ 2011, \mnras, 417, 2809 

%Carbon dioxide-methanol intermolecular complexes in interstellar grain mantles
\bibitem[Dartois et al.(1999)]{Dartois1999} Dartois, E., Demyk, K., d'Hendecourt, L., \& Ehrenfreund, P.\ 1999, \aap, 351, 1066 

%Heavy ion irradiation of crystalline water ice. Cosmic ray amorphisation cross-section and sputtering yield
\bibitem[Dartois et al.(2015)]{Dartois2015} Dartois, E., Aug{\'e}, B., Boduch, P., et al.\ 2015, \aap, 576, A125 

%
\bibitem[Dartois et al.(2018)]{Dartois2018} Dartois, E., Chabot, M., Id Barkach, T., et al.\ 2018, \aap, 618, A173 

%
\bibitem[D'Hendecourt \& Allamandola(1986)]{ldh1986} D'Hendecourt, L.~B., \& Allamandola, L.~J.\ 1986, \aaps, 64, 453 

%Laboratory studies of thermally processed H_2O-CH_3OH-CO_2 ice mixtures and their astrophysical implications
\bibitem[Ehrenfreund et al.(1999)]{Ehrenfreund1999} Ehrenfreund, P., Kerkhof, O., Schutte, W.~A., et al.\ 1999, \aap, 350, 240 

%Energetic processing of laboratory ice analogs: UV photolysis versus ion bombardment
\bibitem[Gerakines et al.(2001)]{Gerakines2001} Gerakines, P.~A., Moore, M.~H., \& Hudson, R.~L.\ 2001, \jgr, 106, 33381 

\bibitem[Gerakines et al.(1995)]{Gerakines1995} Gerakines, P.~A., Schutte, W.~A., Greenberg, J.~M., \& van Dishoeck, E.~F.\ 1995, \aap, 296, 810 

\bibitem[H{\aa}kansson \& Sundqvist(1989)]{Hakansson1989} H{\aa}kansson, P., \& Sundqvist, B.~U.~R.\ 1989, Vacuum, 39, 397 

\bibitem[H{\aa}kansson et al.(1982)]{Hakansson1982} H{\aa}kansson, P., Kamensky, I., \& Sundqvist, B.\ 1982, Nuclear Instruments and Methods in Physics Research, 198, 43 

%Molecular size effects in fast heavy ion induced desorption of biomolecules
%\bibitem[]{Hedin1987}P. HŒkansson , I. Kamensky , M. Salehpour , B. Sundqvist & S. Widdiyasekera (1984) , Radiation Effects, 80:1-2, 141-151

%Fast-ion-induced erosion of leucine as a function of the electronic stopping power
\bibitem[Hedin et al.(1987)]{Hedin1987} Hedin, A., Hkansson, P., Salehpour, M., \& Sundqvist, B.~U.~R.\ 1987, \prb, 35, 7377 

%Absolute infrared intensities and band shapes in pure solid CO and CO in some solid matrices
\bibitem[Jiang et al.(1975)]{Jiang1975} Jiang, G.~J., Person, W.~B., \& Brown, K.~G.\ 1975, \jcp, 62, 1201 

%The Spatial Distribution of Complex Organic Molecules in the L1544 Pre-stellar Core
\bibitem[Jim{\'e}nez-Serra et al.(2016)]{Jimenez-Serra2016} Jim{\'e}nez-Serra, I., Vasyunin, A.~I., Caselli, P., et al.\ 2016, \apjl, 830, L6 

%The ALMA Protostellar Interferometric Line Survey (PILS). First results from an unbiased submillimeter wavelength line survey of the Class 0 protostellar binary IRAS 16293-2422 with ALMA
\bibitem[J{\o}rgensen et al.(2016)]{Jorgensen2016} J{\o}rgensen, J.~K., van der Wiel, M.~H.~D., Coutens, A., et al.\ 2016, \aap, 595, A117 

%
\bibitem[Kaur et al.(2015)]{Kaur2015} Kaur, J., Naghmaa, R.  \& Bobby, A.\ 2015, RSC Adv.,5, 20090

%Molecular complexes theoretical computations between methanol and carbon dioxide and their implications in the interstellar ice mantles
\bibitem[Klotz et al.(2004)]{Klotz2004} Klotz, A., Ward, T., \& Dartois, E.\ 2004, \aap, 416, 801 

%Advances in understanding of swift heavy-ion tracks in complex ceramics
\bibitem[Lang et al.(2015)]{Lang2015} Lang, M., Devanathan, R., Toulemonde, M., \& Trautmann, C.\ 2015, Current Opinion in Solid State and Materials Science, 19, 39 

%Astrochemical evolution along star formation: overview of the IRAM Large Program ASAI
\bibitem[Lefloch et al.(2018)]{Lefloch2018} Lefloch, B., Bachiller, R., Ceccarelli, C., et al.\ 2018, \mnras, 477, 4792 

%Methanol ice co-desorption as a mechanism to explain cold methanol in the gas-phase
\bibitem[Ligterink et al.(2018)]{Ligterink2018} Ligterink, N.~F.~W., Walsh, C., Bhuin, R.~G., et al.\ 2018, \aap, 612, A88 

%Complex organics in IRAS 4A revisited with ALMA and PdBI: Striking contrast between two neighbouring protostellar cores
\bibitem[L{\'o}pez-Sepulcre et al.(2017)]{Lopez-Sepulcre2017} L{\'o}pez-Sepulcre, A., Sakai, N., Neri, R., et al.\ 2017, \aap, 606, A121 
%
\bibitem[Mat{\'e} et al.(2009)]{Mate2009} Mat{\'e}, B., G{\'a}lvez, {\'O}., Herrero, V.~J., \& Escribano, R.\ 2009, \apj, 690, 486 
%
%Radiolysis and sputtering of carbon dioxide ice induced by swift Ti, Ni, and Xe ions
\bibitem[Mej{\'{\i}}a et al.(2015)]{Mejia2015} Mej{\'{\i}}a, C., Bender, M., Severin, D., et al.\ 2015, Nuclear Instruments and Methods in Physics Research B, 365, 477 

%Hydrogenation of CO-bearing species on grains: unexpected chemical desorption of CO
\bibitem[Minissale et al.(2016)]{Minissale2016} Minissale, M., Moudens, A., Baouche, S., Chaabouni, H., \& Dulieu, F.\ 2016, \mnras, 458, 2953 

%Complex Organic Molecules during Low-mass Star Formation: Pilot Survey Results
\bibitem[{\"O}berg, Lauck, \& Graninger(2014)]{Oberg2014} {\"O}berg, K.~I., Lauck, T., \& Graninger, D.\ 2014, \apj, 788, 68 

\bibitem[Palumbo et al.(1999)]{Palumbo1999} Palumbo, M.~E., Castorina, A.~C., \& Strazzulla, G.\ 1999, \aap, 342, 551 

%Spectroscopic constraints on CH3OH formation: CO mixed with CH3OH ices towards young stellar objects
\bibitem[Penteado et al.(2015)]{Penteado2015} Penteado, E.~M., Boogert, A.~C.~A., Pontoppidan, K.~M., et al.\ 2015, \mnras, 454, 531 

%UV radiation field inside dense clouds - Its possible existence and chemical implications
\bibitem[Prasad \& Tarafdar(1983)]{Prasad1983} Prasad, S.~S., \& Tarafdar, S.~P.\ 1983, \apj, 267, 603 

%Modification of ices by cosmic rays and solar wind
\bibitem[Rothard et al.(2017)]{Rothard2017} Rothard, H., Domaracka, A., Boduch, P., et al.\ 2017, Journal of Physics B Atomic Molecular Physics, 50, 062001 

%Infrared optical properties of thin CO, NO, CH4, HC1, N2O, O2, AR, and air cryofilms
\bibitem[Roux et al.(1980)]{Roux1980} Roux, J.~A., Wood, B.~E., Smith, A.~M., \& Plyer, R.~R.\ 1980, Final Report, 1 Oct.~1978 - 1 Sep.~1979 ARO, Inc., Arnold Air Force Station, TN.

\bibitem[Sandford \& Allamandola(1993)]{Sandford1993} Sandford, S.~A., \& Allamandola, L.~J.\ 1993, \apj, 417, 815 

%Laboratory simulation of heavy-ion cosmic-ray interaction with condensed CO
\bibitem[Seperuelo Duarte et al. (2010)]{Seperuelo2010} Seperuelo Duarte, E., Domaracka, A., Boduch, P., Rothard, H., Dartois, E., \& da Silveira, E.~F.\ 2010, \aap, 512, A71 

%Cosmic ray induced explosive chemical desorption in dense clouds
\bibitem[Shen et al.(2004)]{Shen2004} Shen, C.~J., Greenberg, J.~M., Schutte, W.~A., \& van Dishoeck, E.~F.\ 2004, \aap, 415, 203 

%Complex Organic Molecules in Taurus Molecular Cloud-1
\bibitem[Soma, Sakai, Watanabe, \& Yamamoto(2018)]{Soma2018} Soma, T., Sakai, N., Watanabe, Y., \& Yamamoto, S.\ 2018, \apj, 854, 116 

%Amorphous track formation in SiO 2
\bibitem[Szenes(1997)]{Szenes1997} Szenes, G.\ 1997, Nuclear Instruments and Methods in Physics Research B, 122, 530 

%Discovery of Methyl Acetate and Gauche Ethyl Formate in Orion
\bibitem[Tercero et al.(2013)]{Tercero2013} Tercero, B., Kleiner, I., Cernicharo, J., et al.\ 2013, \apjl, 770, L13 

%Transient thermal processes in heavy ion irradiation of crystalline inorganic insulators
\bibitem[Toulemonde et al.(2000)]{Toulemonde2000} Toulemonde, M., Dufour, C., Meftah, A., \& Paumier, E.\ 2000, Nuclear Instruments and Methods in Physics Research B, 166, 903 

%Organic Species in Infrared Dark Clouds
\bibitem[Vasyunina et al.(2014)]{Vasyunina2014} Vasyunina, T., Vasyunin, A.~I., Herbst, E., 
et al.\ 2014, \apj, 780, 85 

%Computation of the electron impact total ionization cross sections of CnH(2n+1)OH molecules from the threshold to 2keV energy range
\bibitem[Vinodkumar et al.(2011)]{Vinodkumar2011} Vinodkumar, M., Korot, K., \& Vinodkumar, P.~C.\ 2011, International Journal of Mass Spectrometry, 305, 26 


%Observational Constraints on Methanol Production in Interstellar and Preplanetary Ices
\bibitem[Whittet et al.(2011)]{Whittet2011} Whittet, D.~C.~B., Cook, A.~M., Herbst, E., Chiar, J.~E., \& Shenoy, S.~S.\ 2011, \apj, 742, 28 

%SRIM - The stopping and range of ions in matter (2010)
\bibitem[Ziegler et al.(2010)]{Ziegler2010} Ziegler, J.~F., 
Ziegler, M.~D., \& Biersack, J.~P.\ 2010, Nuclear Instruments and Methods in Physics Research B, 268, 1818 

\end{thebibliography}
\end{document}